%% file: stopZpaper.tex
\documentclass[nopacs,twocolumn,epjc3]{svjour3}
\pdfoutput=1

\usepackage{atlasphysics}
\usepackage{subfigure}
\usepackage{mathrsfs}
\usepackage{graphicx}
\usepackage{multirow}
\usepackage{rotating}
\usepackage{url}
\usepackage{color}
\usepackage[numbers, sort&compress]{natbib}
\usepackage{setspace}
\usepackage{amsmath}
\usepackage{placeins}
\usepackage{hyperref}
\hypersetup{
colorlinks=true,
linkcolor=blue,
citecolor=blue,
urlcolor=blue
}

\usepackage{preprintcover}  
\PreprintCoverPaperTitle{Search for direct top squark pair production in events with a $Z$ boson, 
$b$-jets and missing transverse momentum in $\sqrt{s}=8$ TeV $pp$ collisions with the ATLAS detector}  
\PreprintIdNumber{CERN-PH-EP-2014-035}  
\PreprintCoverAbstract{A search is presented for direct top squark pair production using
events with at least two leptons including a same-flavour opposite-sign pair with invariant 
mass consistent with the $Z$ boson mass, jets tagged as originating from $b$-quarks and missing transverse momentum.  
The analysis is performed with proton-proton 
collision data at $\sqrt{s}=8\TeV$ collected with the ATLAS detector at the LHC in 2012 
corresponding to an integrated luminosity of 20.3~fb$^{-1}$. 
No excess beyond the Standard Model expectation is observed.
Interpretations of the results are provided in models based on the direct 
pair production of the heavier top squark state ($\tilde{t}_2$) followed by the decay  
to the lighter top squark state ($\tilde{t}_1$) via $\tilde{t}_2 \to Z\tilde{t}_1$, and
for $\tilde{t}_1$ pair production in natural gauge-mediated supersymmetry breaking scenarios where the neutralino 
($\tilde{\chi}^0_1$) is the next-to-lightest supersymmetric
particle and decays producing a $Z$ boson and a gravitino ($\tilde{G}$) via the $\tilde{\chi}^0_1 \to Z\tilde{G}$ process.} 
\PreprintJournalName{EPJC}  

\begin{document}

\titlerunning{Search for direct top squark pair production in events with a $Z$ boson, 
$b$-jets and and $\met$ at ATLAS}

\title{
Search for direct top squark pair production in events with a $Z$ boson, 
$b$-jets and missing transverse momentum in $\sqrt{s}=8$ TeV $pp$ collisions with the ATLAS detector
}

\author{The ATLAS Collaboration}

\date{}

\maketitle

\begin{abstract}

A search is presented for direct top squark pair production using
events with at least two leptons including a same-flavour opposite-sign pair with invariant 
mass consistent with the $Z$ boson mass, jets tagged as originating from $b$-quarks and missing transverse momentum.  
The analysis is performed with proton-proton 
collision data at $\sqrt{s}=8\TeV$ collected with the ATLAS detector at the LHC in 2012 
corresponding to an integrated luminosity of 20.3~fb$^{-1}$. 
No excess beyond the Standard Model expectation is observed.
Interpretations of the results are provided in models based on the direct 
pair production of the heavier top squark state ($\tilde{t}_2$) followed by the decay 
to the lighter top squark state ($\tilde{t}_1$) via $\tilde{t}_2 \to Z\tilde{t}_1$, and
for $\tilde{t}_1$ pair production in natural gauge-mediated supersymmetry breaking scenarios where the neutralino 
($\tilde{\chi}^0_1$) is the next-to-lightest supersymmetric
particle and decays producing a $Z$ boson and a gravitino ($\tilde{G}$) via the $\tilde{\chi}^0_1 \to Z\tilde{G}$ process. 

\end{abstract}

\newcommand{\pho}{\phantom{0}}
\newcommand{\bslash}{\ensuremath{\backslash}}
\newcommand{\BibTeX}{{\sc Bib\TeX}}
\newcommand{\tone}   {\tilde{t}^{}_{1}}
\newcommand{\tonebar}   {\tilde{t}^{*}_{1}}
\newcommand{\ttwo}   {\tilde{t}^{}_{2}}
\newcommand{\ttwobar}   {\tilde{t}^{*}_{2}}

\section{Introduction}
\label{Introduction}
Supersymmetry 
(SUSY)~\cite{Miyazawa:1966,Ramond:1971gb,Golfand:1971iw,Neveu:1971rx,Neveu:1971iv,Gervais:1971ji,Volkov:1973ix,Wess:1973kz,Wess:1974tw} is 
an extension of the Standard Model (SM) which predicts new bosonic partners for the existing fermions and 
fermionic partners for the known bosons. 
In the framework of a generic $R$-parity conserving minimal supersymmetric extension of 
the SM (MSSM)~\cite{Fayet:1976et,Fayet:1977yc,Farrar:1978xj,Fayet:1979sa,Dimopoulos:1981zb}, SUSY particles are produced 
in pairs and the lightest supersymmetric particle (LSP) is stable, providing a possible dark matter candidate.

In a large variety of models, the LSP is the lightest neutralino ($\tilde{\chi}_{1}^{0}$) which is a mixture 
of the neutral supersymmetric partners of the gauge and Higgs bosons, 
known as gauginos and higgsinos. Similarly, charginos are a mixture of the charged gauginos and higgsinos, 
with the lightest denoted by $\tilde{\chi}_{1}^{\pm}$.
The scalar partners of right-handed and left-handed quarks, $\tilde{q}_R$ and $\tilde{q}_L$, mix to form two mass 
eigenstates, $\tilde{q}_1$ and $\tilde{q}_2$, with $\tilde{q}_1$ defined to be the lighter of the two. 
Naturalness arguments~\cite{Barbieri:1987fn,deCarlos:1993yy} imply that the supersymmetric partners of the top quark (stops) are light, 
with mass below $1\TeV$.

Searches for direct pair production of the $\tilde{t}_1$ have been 
performed by the ATLAS~\cite{AtlasStop7TeV,AtlasStop2lep7TeV,AtlasSbottom8TeV,AtlasStop0lep,AtlasStop1lep,AtlasStop2lep} and CMS~\cite{Chatrchyan:2012wa,CMS_SS_bjet,CMS_alphaT_bjet,CMS_stop_1lep} collaborations. 
These searches with $\tilde{t}_{1}\to t\tilde{\chi}_{1}^{0}$ currently have little sensitivity to scenarios
where the lightest stop is only slightly heavier than the sum of the masses of the top quark
and the LSP, due to the similarities in kinematics with SM top pair production ($\ttbar$). 
In those scenarios, by considering instead the direct pair production of the heavy stop ($\ttwo$) decaying via 
$\tilde{t}_{2}\to Z \tilde{t}_{1}$, 
stop signals can be discriminated from the $\ttbar$ background 
by requiring a same-flavour 
opposite-sign (SFOS) lepton pair originating from the $Z$ boson decay. 
Requiring a third lepton, that in signal events can be produced from the top quark in the $\tilde{t}_{1}\to t\tilde{\chi}_{1}^{0}$ decay, can further reject $\ttbar$.
Sensitivity to direct $\tilde{t}_{2}$ pair production can be obtained with this three-lepton signature even in models 
where additional decay modes of the $\tilde{t}_{2}$, such as $\tilde{t}_{2}\to  t\tilde{\chi}_{1}^{0}$ or via the lightest Higgs boson ($h$) in $\tilde{t}_{2}\to h\tilde{t}_{1}$, are significant.

A similar signature can also occur in $\tone$ pair production in gauge-mediated SUSY breaking (GMSB) models~\cite{Dine:1981gu,AlvarezGaume:1981wy,Nappi:1982hm,Dine:1993yw, Dine:1994vc,Dine:1995ag}.
The $\tilde{\chi}_{1}^{0}$ from $\tilde{t}_{1}$ decay is typically the next-to-lightest supersymmetric particle (NLSP) and  the supersymmetric partner of the graviton (gravitino, $\tilde{G}$) 
is typically the LSP and is very light ($m_{\tilde{G}} < 1 \keV$). Assuming a mass scale of the messengers responsible for 
the supersymmetry breaking of around $10\TeV$ and little fine tuning~\cite{Barbieri:1987fn}, the lightest stop is expected to have a mass of less than 
$400\GeV$~\cite{Asano_Naturalness}.
The $\tilde{\chi}_{1}^{0}$ decays to either a $\gamma$, $Z$, or $h$ boson and a $\tilde{G}$. If the $\tilde{\chi}_{1}^{0}$ is higgsino-like,
as suggested by naturalness arguments,
it dominantly decays either via $\tilde{\chi}_{1}^{0}\rightarrow h\tilde{G}$ or via $\tilde{\chi}_{1}^{0}\rightarrow Z\tilde{G}$, 
in the latter case giving a $Z$ boson at the end of the stop decay chain.

In this paper a search for stop pair production is reported in final states characterised by the presence of a $Z$ boson 
with or without additional leptons, 
plus jets originating from $b$-quarks ($b$-jets) produced in the stop decay chain
 and significant missing transverse momentum from the undetected LSPs. 
Results are interpreted in simplified models featuring $\tilde{t}_2$ production and in the framework of natural GMSB.
This paper presents the first result on $\tilde{t}_2$ direct pair production and extends the results of a previous ATLAS analysis, carried out using $7\TeV$ data corresponding to an integrated luminosity of 2.05~\ifb~\cite{AtlasStopGMSB2011}, that excluded stop masses up to $310\GeV$
for $115\GeV<m_{\tilde{\chi}_{1}^{0}}<230\GeV$ in natural GMSB scenarios.

\section{The ATLAS detector}

ATLAS~\cite{AtlasDetector} is a general-purpose particle physics experiment at the LHC. 
The layout of the detector consists of inner tracking devices surrounded by a superconducting solenoid, electromagnetic 
and hadronic calorimeters and a muon spectrometer with a magnetic field produced by three large superconducting toroids 
each with eight coils.
The inner tracking detector is formed from silicon pixel and microstrip detectors, and a
straw tube transition radiation tracker, and provides precision 
tracking of charged particles for pseudorapidity $|\eta|<2.5$
\footnote{ATLAS uses a right-handed coordinate system with its
origin at the nominal {\it pp} interaction point (IP) in the center of the
detector and the $z$-axis along the beam. The $x$-axis points
from the IP to the center of the LHC ring, and the $y$-axis points
upward. ~ Cylindrical coordinates $(r,\phi)$ are used in the transverse
plane, $\phi$ being the azimuthal angle around the $z$-axis. ~ The
pseudorapidity is defined in terms of the polar angle $\theta$ as
$\eta=-\ln\tan(\theta/2)$.  ~ The separation between final state
particles is defined as $\Delta R=\sqrt{(\Delta \eta)^2+(\Delta\phi)^2}$.
~ The transverse momentum is denoted as \pt.}.
The calorimeter system, placed outside the solenoid, covers $|\eta|<4.9$ 
and is composed of electromagnetic and hadronic sampling calorimeters 
with either liquid argon or scintillating tiles as the active medium. 
The muon spectrometer surrounds the calorimeter and consists of a 
system of precision tracking chambers with\-in $|\eta|<2.7$, and 
detectors for triggering within $|\eta|<2.4$.

\section{Signal and background simulation}
\label{MonteCarlo}

Monte Carlo (MC) simulated event samples are used to aid in the
estimation of the SM background and to model the SUSY signal. 
MC samples are processed through a detector simulation~\cite{AtlasSimulation} based on 
{\sc Geant4}~\cite{GEANT4} 
or a fast simulation using a parameterisation of the performance of the electromagnetic and hadronic calorimeters and {\sc Geant4} 
for the other parts of the detector~\cite{FastCaloSim},   
and are reconstructed in the same manner as the data. 
The simulation includes the effect of multiple $pp$ collisions in the same and neighbouring bunch crossings and is 
weighted to reproduce the observed distribution of the average number of 
collisions per bunch crossing. 
All MC samples used in the analysis are produced using the ATLAS underlying e\-vent tune 2B~\cite{mc11ctunes} unless otherwise stated.

The top-quark pair production background is simulated with {\sc Powheg Box} r2129~\cite{Nason:2004rx,POWHEG2,POWHEG3}  interfaced to 
{\sc Pythia 6.427}~\cite{pythia} for the fragmentation and hadronisation processes. The mass of the top quark is fixed 
at 172.5~\GeV, and the next-to-leading order (NLO) parton distribution
function (PDF) set CT10~\cite{Nadolsky:2008zw} is used. The total cross section is calculated at next-to-next-to-leading-order (NNLO) 
including resummation of next-to-next-to-leading logarithmic (NNLL) soft gluon terms with {\sc top++2.0}~\cite{Cacciari:2011hy,Baernreuther:2012ws,Czakon:2012zr,Czakon:2012pz,Czakon:2013goa,Czakon:2011xx}.
The P2011C~\cite{P2011C} MC tune is used for this sample. 
Samples generated with {\sc Alpgen 2.14}~\cite{ALPGEN} 
interfaced with {\sc Herwig 6.510}~\cite{herwig}, including {\sc Jimmy 4.3}~\cite{JIMMY} for the underlying event description, are used to evaluate generator systematic uncertainties,
while {\sc Powheg Box} r2129 interfaced to {\sc Herwig 6.510} and {\sc AcerMC 3.8}~\cite{Kersevan:2013} interfaced to {\sc Pythia 6.426} are
used for hadronisation and initial/final state radiation (ISR/FSR) uncertainty estimation respectively.
Production of a single top quark in association with a $W$ boson is simulated with {\sc Powheg Box} r2129 interfaced to {\sc Pythia 6.426} 
using the diagram removal scheme~\cite{Wt_Powheg_interference}.
The nominal samples describing $\ttbar$ production in association with gauge bosons ($\ttbar V$) 
as well as single top production in association with a $Z$ boson ($tZ$)
 in the $t$- and $s$-channels, and the $tWZ$ process, are generated using 
the leading-order (LO) generator {\sc MadGraph5 1.3.33}~\cite{MadGraph} interfaced to {\sc Pythia 6.426} 
for the fragmentation and the hadronisation. 
The total cross sections of $\ttbar W$ and $\ttbar Z$ are normalised to NLO~\cite{ttbarV-Xsection} 
while $tZ$ is normalized to the LO cross section from the generator, since NLO calculations are currently only available for the $t$-channel~\cite{Campbell:2013yla}. 
To estimate generator and hadronisation systematic uncertainties for the $\ttbar W$ and $\ttbar Z$ processes, {\sc Alpgen 2.14} interfaced with {\sc Herwig 6.520}, 
including {\sc Jimmy 4.3}, is used.
Samples of $Z/\gamma^*$ production in association with up to five jets are produced with {\sc Sherpa 1.4.1}~\cite{Gleisberg:2008ta} where  $b$- and $c$-quarks are treated as massive.
MC samples of dibosons ($ZZ$, $WZ$ and $WW$) decaying to final states with 2, 3 and 4 leptons are generated using 
{\sc Powheg Box} r2129 interfaced to {\sc Pythia 8.163}~\cite{PYTHIA8}. Samples generated with a{\sc MC@NLO}~\cite{aMCAtNLO_diboson} (in {\sc MadGraph5 2.0.0}.beta)
 interfaced to {\sc Pythia 6.427} or {\sc Herwig 6.510} are
used to evaluate generator, hadronisation and scale variation uncertainties.
Samples of tribosons ($WWW$, $ZWW$ and $ZZZ$) 
are generated
with {\sc MadGraph5 1.3.33} interfaced to  {\sc Pythia 6.426} and normalized to NLO~\cite{Campanario:2008yg}. 
Higgs boson production in association with a vector boson or $\ttbar$ pair is simulated with {\sc Pythia 8.165},
with cross sections calculated at NNLO QCD + NLO electroweak precision, except $pp\to\ttbar h$, 
which is calculated at NLO QCD precision~\cite{Dittmaier:2012vm}.
The multijet and $\gamma$+jet processes are simulated with {\sc Pythia 8.165} and {\sc Pythia 8.160} respectively. 

Signal events are generated according to SUSY models using  {\sc Herwig++ 2.5.2}~\cite{Bahr:2008pv} with the CTEQ6L1 PDF set.
 Signal cross sections are calculated 
at NLO + NLL accuracy~\cite{Beenakker:1997ut,Beenakker:2010nq,Beenakker:2011fu}. The nominal cross section and the uncertainty
 are taken from an envelope of cross section predictions using different PDF sets and factorisation and renormalisation scales, as described in Ref.~\cite{Kramer:2012bx}. 
 
Direct $\ttwo$ pair production is studied using a simplified model, where all SUSY particles are decoupled except for the $\tilde{t}_{2}$, $\tilde{t}_{1}$ and $\tilde{\chi}_{1}^{0}$, assumed to be the LSP. The only decays included in this model are $\ttwo\to Z\tone$ and $\tone \to t\tilde{\chi}_1^{0}$. 
The mass of the top quark is fixed at 172.5~\GeV. 
The mass difference between the 
lighter stop and the neutralino is set to $180\GeV$, a region not excluded by previous searches~\cite{AtlasStop1lep}, and 
signal samples are generated varying the masses of the $\ttwo$ and $\tilde{\chi}_1^{0}$. 
In addition, dedicated samples also including the $\ttwo\to h\tone$ and $\ttwo \to t\tilde{\chi}_1^{0}$ decay modes are used 
to interpret the results as a function of the $\ttwo$ branching ratios.
Simulated samples corresponding to direct $\tone$ pair 
production for values of $m_{\tone}=m_{\tilde{\chi}_1^{0}}+180\GeV$ are also used in the analysis.

For the natural GMSB scenario, a very similar model to that of Ref.~\cite{AtlasStopGMSB2011} is considered, with the Higgs boson assumed to be SM-like and with the mass set at $126\GeV$, 
in agreement with the observation of a Higgs boson at the LHC~\cite{ATLAS_Higgs_discovery,CMS_Higgs_discovery}, and with $\tan \beta $, the ratio of the vacuum expectation value of the 
two neutral Higgs doublets of the MSSM, set to 5. 
The masses of the first and second generation squarks and gluinos (superpartners of the gluons) are above $5\TeV$, 
and maximal mixing between the squark eigenstates is assumed for $\tilde{t}_{1}$. Only $\tilde{t}_{1}$ pair production is considered.
$\tilde{\chi}_{1}^{0}$, $\tilde{\chi}_{2}^{0}$ and
$\tilde{\chi}_{1}^{\pm}$ are assumed to be predominantly higgsino states. Hence, if $\tilde{\chi}_{2}^{0}$ or $\tilde{\chi}_{1}^{\pm}$ are produced in a decay chain,
they decay to $\tilde{\chi}_{1}^{0}$ promptly with soft accompanying fermions. The branching fractions of the $\tilde{t}_{1}$ and higgsino decays are predicted by the model. 
If $m_{\tilde{t}_{1}}<m_{t}+m_{\tilde{\chi}_{1}^{0}}$, $\tilde{t}_{1}$ decays via $\tilde{t}_{1}\rightarrow b\tilde{\chi}_{1}^{\pm}$ exclusively, while if 
$m_{\tilde{t}_{1}}>m_{t}+m_{\tilde{\chi}_{1}^{0}}$, $\tilde{t}_{1}$ may also decay with similar probability via $\tilde{t}_{1}\rightarrow t\tilde{\chi}_{1}^{0}$ (or $t\tilde{\chi}_{2}^{0}$). 
For the model parameters considered, the $\tilde{\chi}_{1}^{0}$ predominantly decays to $Z\tilde{G}$ with branching ratios typically above 70\%. 
Signal samples are generated varying the $\tilde{t}_1$ and $\tilde{\chi}_1^{0}$ masses.

\section{Object identification and event selection}
\label{sec:obj}

After the application of beam, detector and data quality requirements,
the total luminosity considered in this analysis corresponds to~20.3~fb$^{-1}$.
The uncertainty on the integrated luminosity is $\pm$2.8\%. It is derived, following the same methodology 
as that detailed in Ref.~\cite{ATLAS_Luminosity2012}, from a preliminary calibration of the luminosity scale 
derived from beam-separation scans performed in No\-vem\-ber 2012.

Events are selected if they pass the single electron or muon triggers; these are fully efficient for lepton $\pt>25\GeV$.
The presence of at least one primary vertex, with at least five tracks with $\pt>0.4\GeV$
associated to it, is required. 
In order to optimize the analysis and to perform data-driven background estimations, two categories of jets, electrons, muons and photons are 
defined: ``candidate'' and ``signal'' (with tighter selection criteria).

Jets are reconstructed from three-dimensional ca\-lorime\-ter energy
clusters by using the anti-$k_t$ algorithm \cite{antikt_algorithm} with a radius parameter of
0.4. Jet energies are corrected \cite{ATLAS_JES_paper} for detector
inhomogeneities, the non-com\-pen\-sa\-ting nature of the calorimeter, and
the impact of multiple overlapping {\it pp} interactions, using
factors derived from test beam, cosmic ray and {\it pp} collision data
and from a detailed {\sc Geant4} detector
simulation. 
Events with any jet that fails the jet quality criteria designed to remove noise and non-collision backgrounds~\cite{ATLAS_JES_paper} are rejected. 
Jet candidates are required to have $\pt>20\GeV$
and $|\eta|<2.8$. 
Jets labelled as signal jets are further required to have $\pt>30\GeV$ 
and, for those with $\pt<50\GeV$ and $|\eta|<2.4$, 
the jet vertex fraction, defined as the fraction of the sum of the  $\pt$
of the tracks associated with the jet and matched to the selected
primary vertex, normalised by the sum of the  $\pt$ of all tracks
associated with the jet, is required to be larger than 25\%.

Identification of jets containing $b$-quarks ($b$-tagging) is performed with a 
dedicated algorithm 
based on a neural-network approach 
which uses the output weights of several $b$-tagging algorithms \cite{BTagHighPerform} as input. 
A requirement is chosen corresponding to a 60\% average efficiency 
obtained for $b$-jets in simulated $\ttbar$ events.
The rejection factors for mis-tagging light quark jets, $c$-quark jets and $\tau$ leptons  in simulated SM $\ttbar$ events 
are approximately 600, 8 and 24, respectively. 
Signal jets with $|\eta|<2.5$ which satisfy this {\it b-}tagging requirement are identified as {\it b-}jets. 
To compensate for differences between data and MC simulation in the $b$-tagging efficiencies and mis-tag rates, 
correction factors derived from different me\-thods, 
such as the use of the $\pt$ of muons relative to the axis of the jet~\cite{ATLAS-CONF-2012-043} 
and a dedicated study in $\ttbar$ dominated regions~\cite{ATLAS-CONF-2012-097}, are applied to the simulated samples. 
A sample of $D^{*+}$ mesons is used for mis-tag rates of {\it c-}jets~\cite{ATLAS-CONF-2012-039} and inclusive jet samples for mis-tag rates of a jet 
which does not originate from a {\it b-} or {\it c-}quark~\cite{ATLAS-CONF-2012-040}.

Electron candidates must satisfy the ``me\-dium'' selection
criteria described in Ref. \cite{ATLAS:electron_performance}, re-op\-ti\-mi\-sed for 2012 data, and are
required to fulfil $\pt > 10\GeV$ and $\vert\eta\vert < 2.47$. Signal
electrons must pass the previous requirements and also need to be isolated, i.e. the scalar sum of the $\pT$ of charged-particle tracks
within a cone of radius $\Delta R= 0.3$ around the candidate excluding its own track must be less than 16\% of the electron $\pT$. In addition, a longitudinal impact parameter requirement of $\vert z_0 \sin\theta\vert <0.4~\rm{mm}$ is applied to signal electrons. 
The track parameter $z_{0}$ is defined with respect to the reconstructed primary vertex.

Muon candidates are required to have $\pT > 10\GeV$, $\vert\eta\vert < 2.4$ and are identified by matching an extrapolated inner detector track and one or more track segments in the muon spectrometer~\cite{ATLAS-CONF-2013-088}. 
Signal muons are then required to be isolated, i.e. the scalar sum of the $\pT$ of charged-particle tracks within a cone of radius $\Delta R = 0.3$ around the muon candidate excluding its own track must be less than 12\% of the muon $\pT$.  In addition, a longitudinal impact parameter requirement of 
$\vert z_0 \sin\theta \vert<0.4~\rm{mm}$ is applied to signal muons. 

A signal lepton with $\pt$ larger than $25\GeV$ is required to match the one that triggered the event 
such that the efficiency of the trigger is $\pt$  independent. 
The MC events are corrected to account for minor differences in the lepton trigger, reconstruction and identification efficiencies 
between data and MC simulation~\cite{ATLAS:electron_performance,ATLAS-CONF-2013-088}. 

\begin{table*}[!hbt]
  \caption{Summary of the event selection in the signal and $t\bar{t}$ background control regions used in the analysis. 
  The variables used are the number of leptons ($N^{\mathrm{leptons}}$), the $\pt$ of the leading lepton ($\pt(\ell_1)$), 
  the dilepton flavour (SF: same-flavour; DF: different flavour), 
  the dilepton invariant mass ($m_{\ell\ell}$), the number of $b$-jets ($N^{b\text{-jets}}$), the number of jets regardless of their flavour ($N^{\rm{jets}}$), the $\pt$ of the leading jet ($\pt({\rm jet}_1)$), the $\pt$ of the $N^{\rm{jets}}$-th jet required in each region ($\pt({\rm jet}_N)$), 
  the missing transverse momentum ($\met$), the transverse momentum of the dilepton system ($\pt({\ell\ell})$), 
  and the angular separation in the 
  transverse plane between the leptons forming the SFOS pair ($\Delta\phi^{\ell\ell}$). 
  }
  \label{tab:SRs}
  \begin{center}
    \begin{tabular}{lccccccc}
      \hline\noalign{\smallskip}
      & SR2A & SR2B & SR2C & CR2A & CR2C & SR3A & SR3B \\
      \hline\noalign{\smallskip}
      $N^{\mathrm{leptons}}$ & 2 & 2 & 2 & 2 & 2  & 3 & 3\\ 
      $\pt(\ell_1)$  [\GeV] &  $>25$ & $>25$ & $>25$  & $>25$ & $>25$ & $>40$  & $>60$ \\ 
      dilepton flavour & SF & SF & SF & SF,DF & SF,DF & SF & SF \\
      $|m_{\ell\ell}-m_Z |$ [\GeV] & $<5$ & $<10$ & $<5$ & $<50$ & $<50$ & $<10$ & $<10$\\ 
      &  &  &  & $>10$ (SF) & $>10$ (SF) & & \\ 
      $N^{b\text{-jets}}$& $\geq 1$ & $\geq 1$ & $\geq 1$ & $\geq 1$ & $\geq 1$ & $\geq 1$ & $\geq 1$\\
      $N^{\rm{jets}}$ & $3$, $4$ & $3$, $4$ & $\geq 5$ & $ 3$, $4$ & $\geq 5$ &  $\geq 5$ & $\geq 5$\\
      $\pt({\rm jet}_1)$ [\GeV] &  $>30$  & $>30$  & $>30$   &  $>30$  & $>30$    & $>50$  & $>40$ \\
      $\pt({\rm jet}_N)$ [\GeV] &  $>30$  & $>30$  &$>30$  &  $>30$  & $>30$  & $>30$  & $>40$ \\
      \met  [GeV] & $>160$ & $>200$ & $>160$  & $>160$  & $>120$   &   $>60$ & $>60$ \\    
      $\pt({\ell\ell})$  [\GeV] & $>80$  & $>160$ & $>80$ &  $>80$  & $>80$ & - & $>75$ \\ 
      $\Delta\phi^{\ell\ell}$ [rad]& $<1.5$ & $<1.5$ & $<1.5$ & $<1.5$ & $<1.5$ & - & -\\
      \hline\noalign{\smallskip}
    \end{tabular}
  \end{center}
\end{table*}

To resolve ambiguities between reconstructed jets and leptons, jet candidates
within a distance of $\Delta R=0.2$ of an electron candidate are rejected. Any electron or muon candidate within a distance of
$\Delta R=0.4$ of any remaining jet candidate is also rejected.
To suppress the rare case where two distinct tracks are
mistakenly associated with one calorimeter energy cluster forming two electron candidates, if two 
electron candidates are found within a distance $\Delta R=  0.1$, the one with smaller transverse momentum 
is rejected. 
Finally, to suppress muon bremsstrahlung 
leading to an incorrect measurement of the transverse momentum,
if an electron candidate and a muon candidate are within $\Delta
R=0.1$, both are rejected.

Photons are used only for the $Z$+jets estimation in the two-lepton signal regions described in Section~\ref{section:background_estimation}
and the overlap removal between photons and jets described below is performed only in this case.
Photon candidates are required to have $\pt>25\GeV$, $|\eta|<2.47$ and must satisfy the ``tight''
selection criteria described in Ref.~\cite{ATLAS_photon_measurement}. Signal photons are further required to be
isolated, i.e. the scalar sum of transverse energy deposition in the calorimeter observed within a cone 
of radius $\Delta R=0.4$ around the photon candidate excluding
its own energy deposition in the calorimeter must be less than $4\GeV$.
To resolve overlaps between reconstructed jets and photons, 
jet candidates within a distance of $\Delta R=0.2$ of a photon candidate
are rejected.

The calculation of the missing transverse momentum, where its magnitude is referred to as $\met$ \cite{ATLAS_MET}, is based on 
the vector sum of the transverse momenta of all electron, muon and jet candidates, as well as photons with $\pt>10$~\GeV\
and calibrated calorimeter energy clusters with $|\eta| < 4.9$
not associated with these objects. Clusters associated with electrons, photons and jets 
make use of the calibrations of these objects. For jets, the calibration includes the pile-up correction described
above, whilst the jet vertex fraction requirement is not considered when selecting jet
candidates for computing the $\met$. Clusters not associated with these objects are calibrated using both
calorimeter and tracker information~\cite{ATLAS_MET}.

Five signal regions (SRs) are defined in the analysis aiming at final states with a $Z$ boson, $b$-jets, significant $\met$ and possibly additional leptons, 
as summarised in Table~\ref{tab:SRs}. 
They are characterised by the number of leptons (electrons or muons) required in the final state.
For the two-lepton SRs (indicated as SR2A, SR2B and SR2C), events with exactly two leptons are selected, 
with the $\pt$ of the leading one required to be larger than $25\GeV$. 
They are required to be signal leptons and form a SFOS pair with 
invariant mass ($m_{\ell\ell}$) within
$5\GeV$ or $10\GeV$ of the $Z$-boson mass.
At least one {\it b-}jet is required.
SR2A and SR2B are optimised for the small $m_{\tilde{t}_{1}}-m_{\tilde{\chi}_{1}^{0}}$ region of
the natural GMSB model where low jet multiplicity is expected,
whilst SR2C is optimised for the large $m_{\tilde{t}_{1}}-m_{\tilde{\chi}_{1}^{0}}$ region where the jet multiplicity is high.
SR2A is optimised for a stop mass around $400\GeV$ and SR2B is for $600\GeV$.
Since the $Z$ boson produced in stop signal events is typically boosted, the transverse momentum of the dilepton system, $\pt(\ell\ell)$,
tends to be high while the azimuthal separation $\Delta\phi^{\ell\ell}$ tends to be low. 
This is illustrated by Figure~\ref{figure:ptll_2lep_plot}, which shows the $\pt(\ell\ell)$ distribution after the lepton, $m_{\ell\ell}$, jet 
and $b$-jet requirements in SR2A are applied. Requirements of  $\Delta\phi^{\ell\ell}$ below 1.5
and $\pt(\ell\ell)>80\GeV$ or $160\GeV$ are therefore applied in the SRs. 
Finally, to enhance the signal contribution, typically with large $\met$ due to the LSPs, $\met>160 \GeV$ or $200 \GeV$ is required depending on the targeted 
stop mass.

\begin{figure}[!hbt]
  \begin{center}
    \raisebox{10pt}{
    \hspace{-0.5cm}
    \includegraphics[scale=0.38]{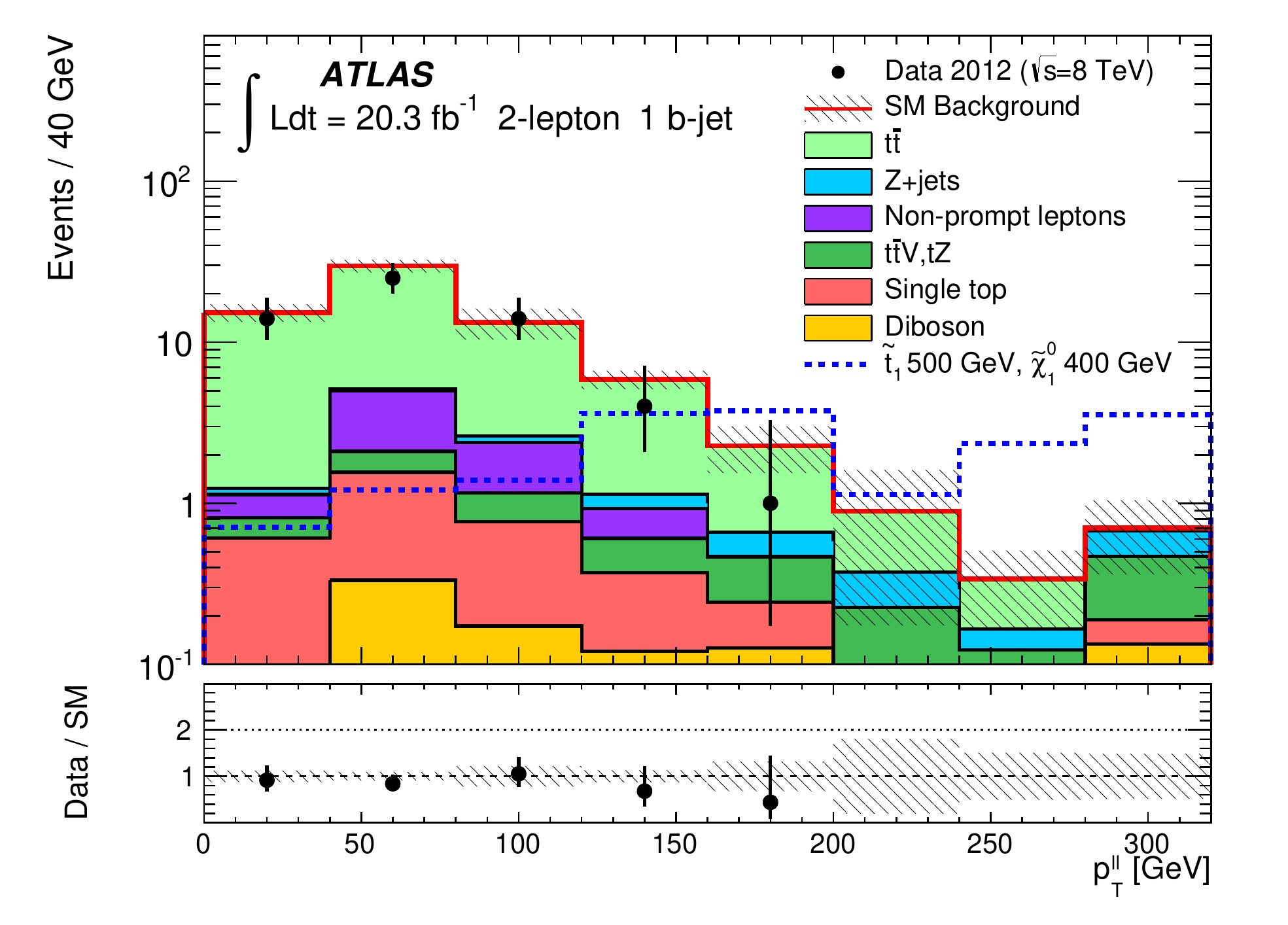}}
  \vspace{-0.2cm}
   \includegraphics[scale=0.412]{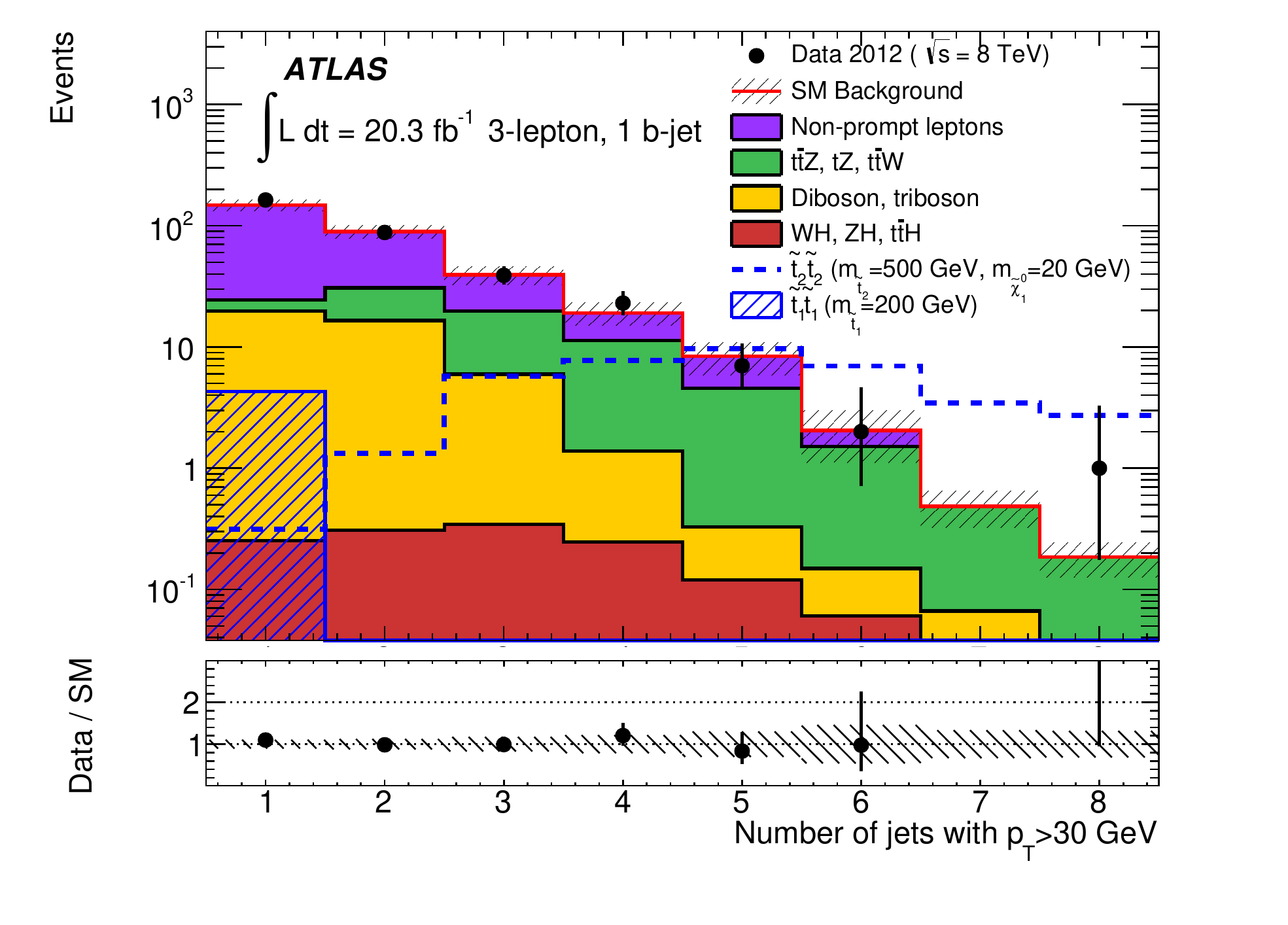}\\
  \end{center}
  \vspace{-0.5cm}
  \caption{{\it Top}, $\pt(\ell\ell)$ distributions in SR2A before the $\pt(\ell\ell)>80$~GeV and $\Delta\phi^{\ell\ell}<1.5$ selections. 
    {\it Bottom}, number of signal jets with $\pt>30$~GeV in events
    with 3 signal leptons after the lepton, $m_{\ell\ell}$ and $b$-jets selections in SR3A. 
    Shaded bands denote the background
    statistical and systematic uncertainty. For illustration, distributions for selected signal points are also
    shown: the stop natural GMSB model with
    $m_{\tilde{t}_{1}}=500\GeV$, $m_{\tilde{\chi}_{1}^{0}}=400\GeV$ ({\it top}) and the simplified model with
    $m_{\tilde{t}_{2}}=500\GeV$, $m_{\tilde{t}_{1}}=200$~GeV and $m_{\tilde{\chi}^{0}_{1}}=20\GeV$ for both direct $\ttwo$ and $\tone$ pair production ({\it bottom}).
    The last bin includes the histogram overflow.
}
  \label{figure:ptll_2lep_plot}
\end{figure}

In the three-lepton SRs (indicated as SR3A and SR3B), at least three signal leptons with
two of them forming an SFOS pair with invariant mass 
which is within $10\GeV$ of the $Z$ boson mass are required.
Two regions are optimised to give good sensitivity in the direct $\ttwo$ pair production model for different $\ttwo - \tone$ mass splittings. 
The SR3A is aimed at signal models with low mass splitting where the $Z$-boson is not boosted.
The SR3B is optimised for high mass splitting where the $Z$-boson is boosted
requiring a minimum $\pt$ of the dilepton system of 75~\GeV. 
A high-$\pt$ leading lepton with a minimum $\pt$ requirement of $40 \GeV$ or $60 \GeV$ for SR3A and SR3B respectively, and at least one {\it b-}jet 
are required  to suppress the diboson background. 
The signal is expected to have higher jet multiplicity than the SM
background, due to the presence of two top quarks and two $Z$ bosons.
This is illustrated by Figure~\ref{figure:ptll_2lep_plot}, which shows the jet multiplicity distribution after the lepton, $m_{\ell\ell}$, 
and $b$-jet requirements in SR3A are applied. 
Therefore at least five jets are required to increase the signal sensitivity.

\section{Background estimation}
\label{section:background_estimation}

Two main sources of background can be distinguished in this analysis: events containing at least one non-prompt or fake lepton 
(mainly production of multijets and $W$ boson in association with jets 
in the two-lepton SRs, and 
production of top pairs and $Z$ boson in association with jets in the three-lepton SRs)
and events with two or three prompt leptons (mainly $Z$+jets and $\ttbar$ in the two-lepton SRs, and  
$\ttbar V$, $tZ$, diboson and triboson events in the three-lepton SRs). 

\paragraph{Background from fake or non-prompt leptons\\}

“Fake” leptons can originate from a misidentified light flavour quark or
gluon jet (referred to as “light flavour”). “Non-prompt” leptons can
originate from a semileptonic decay of a hadron containing a $b$- or $c$-quark (referred to as “heavy flavour”), or an
electron from a photon conversion.
The contribution from fake and non-prompt leptons is estimated from data with a matrix method similar to that described in 
Refs.~\cite{Top1lep,Top2lep}.
In order to perform the matrix method, two types of lepton identification criteria are defined: ``tight'', corresponding to the signal lepton criteria described in Section~\ref{sec:obj}, and ``loose'', corresponding to candidate leptons. 
To increase the available statistics, muons within a $0.2 < \Delta R < 0.4$ distance from jets are also considered as loose muons in the method if 
the scalar sum of $\pT$ of charged-particle tracks within a cone of radius $\Delta R = 0.3$ 
around the muon candidate excluding its own track is less than 30\% of the muon $\pT$. 
The matrix method relates the number of events containing fake or non-prompt leptons to the number of observed events with tight or loose leptons 
using the probability for loose prompt, fake or non-prompt leptons to pass the tight criteria. 
The probability for loose prompt leptons to pass the tight selection criteria is obtained using a $Z\to\ell\ell$ data sample. 
The probability for loose non-prompt leptons to pass the tight selection criteria is determined from data 
separately for heavy flavour in a $\bbbar$ enriched sample and for photon conversions in a $Z\to\mu\mu\gamma$ sample. 
This probability is modelled as a function of $\pt$ and $\eta$ for electrons
and of $\pt$ and the number of jets for muons.  
Simulation studies show that the contribution of fake leptons originating from 
a misidentified light flavour quark or gluon jet is negligible in all the signal and 
data control regions used for the background estimation. 
The probability for loose non-prompt electrons passing the tight selection is calculated according to the fraction of  
heavy flavour and photon conversion obtained in MC for the different regions.

For SRs with two leptons, relations 
are obtained  for the observed event counts 
as a function of the number of events containing prompt and non-prompt leptons.  
These can be solved simultaneously to estimate the number of background  
events with two tight lepton candidates with at least one non-prompt lepton. 
In the three-lepton SRs, the background from non-prompt leptons is estimated as in the two-lepton case
by considering the leading lepton to be prompt, 
which simulation studies show to be true in $>$99\% of the events,
and applying the same estimation method to the second and third leading leptons in the event. 
The results of the estimations have been validated with data in regions with similar background composition
obtained by reversing the $\met$ or jet multiplicity cuts used in the SRs.

\paragraph{$\ttbar$ background in the two-lepton channel \\}

The dominant background in the two-lepton signal regions comes from $\ttbar$. 
The background prediction is normalized to data in dedicated control regions (CRs),  
and then extrapolated to the SRs. 
The observed number of events in the CRs are used to derive $\ttbar$ estimates in each of the SRs
via a profile likelihood method~\cite{AsymptoticFormula}. 

The CRs are designed to have kinematic selections as similar as possible to the corresponding SRs in order to minimize systematic 
uncertainties on the extrapolation of the background to the SR. 
The CRs use both dilepton events with the same flavour (SF) and different flavour (DF) with the following dilepton mass requirements: 
$10\GeV<|m_{\ell\ell}-m_{Z}|<50\GeV$ (SF), and $|m_{\ell\ell}-m_{Z}|<50\GeV$ (DF). 
Except for lepton-flavour dependent systematic uncertainties, SF and DF events are treated in the same way.
Apart from the $m_{\ell\ell}$ requirements the CR corresponding to SR2A/B (labelled CR2A) has exactly the same selections as SR2A, whereas the CR for SR2C (labelled CR2C) 
has a looser $\met$ selection than the SR to increase the number of events in the CR.

For the background estimation neglecting any possible signal contribution in the CRs, the fit takes as input 
the number of expected background events in each CR and SR 
taken from MC or data-driven estimations  
and the number of observed events in the CRs. 
For each SR, the free parameter is
the overall normalisation of the $\ttbar$ process. 
Each uncertainty source is treated as a nuisance parameter in the fit, constrained with a Gaussian
function taking into account the correlations between different background sources. The likelihood function is the
product of Poisson probability functions describing the observed and expected number of events in the CRs, 
and the Gaussian constraints on the nuisance parameters. 
The contribution from all other non-constrained
processes are set at the theoretical expectation, but are allowed to vary within their uncertainties.
The fitting procedure maximises this likelihood by adjusting the free and nuisance parameters.
For the signal models considered in this paper the contamination of the CRs by signal events is small (typically less than 10\%).

The expected and observed number of events in the control regions are shown in Table~\ref{table:ttbar_CR_num}.
The MC simulation before the fit overestimates the number of $\ttbar$ events observed in both of the CRs. This mis-modelling at 
high $\ttbar$ transverse momentum ($p_{\text{T},\ttbar}$) has been observed in previous ATLAS analyses~\cite{TopDifferentialPaper}.
 
\begin{table}[t]
  \caption{Background fit results and observed numbers of events in the $t\bar{t}$
    control regions for the two-lepton channel.
    The uncertainty shown is the sum of the statistical and systematic uncertainties. Nominal MC expectations are given for comparison.}
  \label{table:ttbar_CR_num}
  \begin{center}
    {\small
    \begin{tabular}{lcc}
      \hline\noalign{\smallskip}
      & CR2A & CR2C \\
      \hline\noalign{\smallskip}
      Data & 152 & 101 \\
      \hline\noalign{\smallskip}
      Fitted total SM & $152\pm 13$ & $101\pm 11$ \\
      \hline\noalign{\smallskip}
      Fitted $t\bar{t}$ &  $128 \pm 13$ & $88\pm 11$ \\
      Fitted single top & $12\pm 4$ & $4.4\pm 3.2$ \\
      Fitted $Z$+jets & $0.62\pm 0.04$ & $0.75\pm 0.07$ \\
      Fitted diboson & $1.6\pm 1.4$ & $0.5 \pm 0.4$ \\
      Fitted $t\bar{t}V,tZ$ & $1.6 \pm 0.4$ & $1.7\pm 0.5$ \\
      Fitted non-prompt & $7.4\pm 2.4$ & $6.1\pm 1.9$ \\
      \hline\noalign{\smallskip}
      MC exp. total SM & 176 & 146 \\
      \hline\noalign{\smallskip}
      MC exp. $t\bar{t}$ & 152 & 132 \\
      MC exp. single top & 13 & 5.2 \\
      MC exp. $Z$+jets & 0.62 & 0.75 \\
      MC exp. diboson & 1.7 & 0.5 \\
      MC exp. $t\bar{t}V,tZ$ & 1.6 & 1.7  \\
      Data-driven non-prompt & 7.4 & 6.1  \\
      \hline\noalign{\smallskip}
    \end{tabular}
  }
\end{center}
\end{table}

\paragraph{$Z$+jets background in the two-lepton channel \\}

Background events from $Z$-boson production associated with jets
typically contain fake $\met$ due to resolution effects in the jet momentum measurement. 
Due to the limited statistics and the difficulty of accurately reproducing fake $\met$ in MC simulations, 
a data-driven ``jet smearing method''~\cite{noleptonSUSY5ifbPub} is used to estimate this
contribution in the high \met\ tail. In this method, well-measured $Z$+jets events
with low $\met$ are selected. By applying jet energy resolution smearing
to these events a pseudo-data sample with fake $\met$ is generated.
The pseudo-data sample is then normalised to data
in the $\met<80\GeV$ region, after subtracting other SM background sources estimated by MC for real two lepton events and 
by the data-driven method for events with non-prompt leptons. Their contribution is less than 10\%. 
The jet energy resolution smearing function ($\pt^{\text{reco}}/\pt^{\text{truth}}$) is initially obtained 
from multijet MC simulation, where $\pt^{\text{reco}}$ is the transverse momentum
of the reconstructed jet and $\pt^{\text{truth}}$ is the transverse momentum of the jet constructed
from stable truth particles excluding muons and neutrinos. Stable particles are defined as
those with a lifetime of 10 ps or more in the laboratory frame.
The function is corrected using $\gamma$+jet data events
where the photon and the jet are balanced. These events are selected by a single photon
trigger and require at least one signal photon and one baseline jet. To suppress soft radiation
that would affect the $\pt$ balance between the jet and the photon, the angle between the leading jet and the leading photon 
in the transverse plane 
is required to be larger than 2.9 rad, and the second-leading jet is required to have $\pt$ of less than
20\% of the $\pt$ of the photon. 
Using the $\pt$ of the balanced photon as reference for that of the jet, the $\pt$ response of jets is measured
in data and MC. The jet energy resolution smearing function is then modified to match $\pt$ response between
data and MC.
The method is validated by closure tests using MC simulation, and also using data in the $80\GeV<\met<160\GeV$ region.

\paragraph{Other backgrounds\\}

The estimation of other background processes producing two or three prompt leptons, such as 
diboson,  triboson, $\ttbar V$, $tZ$ or $Wt$ production,  
is performed using the MC samples described in Section~\ref{MonteCarlo}.

Since $\ttbar Z$ is the main background in the three-lepton SRs and 
has a topology very similar to a $\ttwo\to Z\tone$ signal, dedicated validation regions with
 an enhanced contribution from this background and orthogonal to the SRs are defined to verify the MC prediction in data. 
These regions are defined requiring at least  three leptons 
and the same $m_{\ell\ell}$ and $b$-jet requirements as the SRs.
In order to enhance the $\ttbar Z$ contribution and reduce the possible contamination from signal events, 
the events are required to have
from three to five jets with $\pt>30$~GeV and fewer than five jets with $\pt>50$~GeV. The $\met$ is required to be less than 150~GeV 
except for events with 5 jets with $\pt>30$~GeV where the $\met$ is required to be less than 60~GeV to avoid 
overlaps with the SRs. The third leading lepton is required to have $\pt>20$~GeV to reduce the contribution from non-prompt leptons.  
Two separate validation regions are defined using the $\pt(\ell\ell)$ variable: VR3A with $\pt(\ell\ell)<120$~GeV and VR3B with $\pt(\ell\ell)>120$~GeV. 
The contamination from a potential signal can be large in these validation regions but would typically affect VR3A and VR3B differently depending 
on the $\ttwo$-$\tone$ mass splitting. 
Table~\ref{table:3l_VR3L} shows the expected number of events in these validation regions taken from MC or data-driven estimations 
together with the observed number of events. 
The expected contribution from selected signal models is also shown. 
The $\ttbar Z$ contribution is 40-50\%  of the total expected event count, and a good agreement with data is observed in both regions.

\begin{table}[t]
\caption{Number of events in the VR3A and VR3B $\ttbar Z$ validation regions together with the expectation for some signal points in the $\ttwo$ simplified model. The errors on the backgrounds include both statistical and systematic uncertainties. Only statistical uncertainties are shown for the signal points.}
\label{table:3l_VR3L}
\begin{center}
{\small
\begin{tabular}{lcc}
\hline
           & VR3A&  VR3B \\ 
\hline\noalign{\smallskip}
Data   & 24 & 13 \\\hline\noalign{\smallskip}
Total SM & 19 $\pm$ 5 & 12.1 $\pm$ 3.2  \\\hline\noalign{\smallskip}
MC exp. $\ttbar Z$
  & 7.9 $\pm$ 2.1 & 5.9 $\pm$ 1.6 \\
MC exp. $ tZ $
  & 2.7 $\pm$ 2.7 & 1.5 $\pm$ 1.5 \\
Data-driven non-prompt                       
 & 5.9 $\pm$ 2.9 & 2.7 $\pm$ 1.4  \\ 
MC exp. diboson, triboson   
 & 1.5 $\pm$ 0.5 & 1.9 $\pm$ 0.6 \\
MC exp. $\ttbar W$
  & 0.35 $\pm$ 0.10 & 0.05 $\pm$ 0.02 \\
MC exp.  $Wh$, $Zh$, $\ttbar h$ 
 & 0.3 $\pm$ 0.3 & 0.05 $\pm$ 0.05 \\ 
 \hline\noalign{\smallskip}
($m_{\ttwo},m_{\tilde{\chi}_1^0}$)=(500,20) \GeV 
 & 1.6 $\pm$ 0.6 & 7.5 $\pm$ 1.2 \\
($m_{\ttwo},m_{\tilde{\chi}_1^0}$)=(500,120) \GeV  
 & 3.3 $\pm$ 0.8 & 3.9 $\pm$ 0.8 \\ 
($m_{\ttwo},m_{\tilde{\chi}_1^0}$)=(550,20) \GeV   
 & 0.6 $\pm$ 0.3 & 4.6 $\pm$ 0.7 \\ 
($m_{\ttwo},m_{\tilde{\chi}_1^0}$)=(550,220) \GeV   
 & 2.7 $\pm$ 0.5 & 2.2 $\pm$ 0.5 \\
 \hline\noalign{\smallskip} 
\end{tabular}
}
\end{center}
\end{table}

\section{Systematic uncertainties}

The dominant detector-related systematic effects are due to the jet
energy scale (JES) and resolution (JER) uncertainties, and the
uncertainties on the $b$-tagging efficiency and mistag rates.

The JES uncertainty is derived from a combination of simulation,
test-beam data and in-situ 
measurements~\cite{ATLAS_JES_paper}. Additional terms
accounting for flavour composition, flavour response, 
pile-up and {\it b-}jet scale uncertainties are taken into account. 
These uncertainties sum to 10-20\% of the total number of estimated background events depending on the SR.
JER uncertainties are determined with an in-situ measurement of the jet
response asymmetry in dijet
events~\cite{ATLAS_JER_paper}, and the impact on the SRs ranges between 1-10\%.
Uncertainties associated with the $b$-tagging efficiency and mis-tagging of a $c$- and light-quark jet
are obtained from the same techniques used in the derivation of their correction factors. 
The uncertainty on the expected number of background events in the SR due to $b$-tagging ranges between 4-10\%. 

For the non-prompt lepton background estimation, uncertainties are assigned 
due to the statistical uncertainty on the number of data events with loose and tight leptons and 
due to the MC uncertainty on the relative composition of non-prompt electrons (heavy flavour and conversions).
The uncertainties on the probabilities for loose leptons to pass the tight selections typically range between 10-45\%, are estimated by using alternative samples for their computation, and include possible dependencies on the lepton $\pt$, $\eta$ or jet multiplicity. 
The overall impact of the non-prompt lepton background uncertainties on the expected number of background events 
are below 2\% in the 2-lepton SRs and approximately 15\% in the 3-lepton SRs.

The uncertainties on the MC modelling of background processes are
determined by testing different generators as well as parton
shower and hadronisation models. 
The systematic uncertainties on the modelling of
$\ttbar$+jets,  used only to determine the transfer factors between control and signal regions in the two-lepton case, 
are evaluated by comparing results obtained with the  {\sc Pow\-heg} and {\sc Alpgen} generators.
The hadronisation uncertainty is addressed by comparing
{\sc Powheg} interfaced to  {\sc Pythia6} with  {\sc Powheg} interfaced
to  {\sc Herwig+Jimmy}. 
The uncertainty related to the amount of ISR/FSR is estimated using the predictions
of dedicated {\sc AcerMC} samples generated with different tuning parameters. 
The uncertainties on $t\bar{t}$ are dominated by
these theoretical uncertainties after the fit.
A 22\% cross section uncertainty is assumed for $\ttbar Z$ and $\ttbar W$~\cite{ttbarV-Xsection}. º
The uncertainties on the modelling of $\ttbar V$ are
evaluated by comparing {\sc MadGraph} interfaced to {\sc Pythia6}
with {\sc Al\-pgen}~interfaced with {\sc Herwig+Jimmy}.
The uncertainty assigned on the diboson 
cross sections are $5\%$ for $ZZ$~\cite{xs:dibosonsZZ} and $7\%$ for $WZ$~\cite{xs:dibosonsWZ}. 
For diboson production processes, the uncertainties on the
modelling are evaluated by comparing {\sc Powheg} interfaced to {\sc Pythia8} with the a{\sc MC@NLO} generator interfaced to {\sc Pythia6} and {\sc Herwig+Jimmy} .  
For tribosons, $\ttbar h$ and $tZ$ production processes, which constitute a very small background in all signal regions, 
a $100\%$ uncertainty 
on the cross section is assumed. 
The uncertainties on these processes are large to account for kinematic effects, 
even though the inclusive cross sections are known to better precision.

\section{Results and interpretation}

The number of data events observed in each SR for the
two-lepton and three-lepton analyses is reported in Table~\ref{table:result}
together with the expected SM background contributions. Figures~\ref{figure:2lep_final_met} and~\ref{fig:3lep_met_finalSR} show the $\met$ distributions for data and background expectations for each SR. 
\begin{table}[htb]
  \caption{Observed event counts and predicted numbers of events for each SM background process in the SRs used
    in the analysis. For two-lepton SRs, background fit results
    and nominal MC expectations are given for comparison. The ``non-prompt'' category includes $\ttbar$, single top and $Z$+jets 
    processes for the three-lepton SRs SR3A and SR3B. 
    The $p$-value of the observed events for the background only hypothesis ($p_{0}$) is also shown.
    The value of $p_0$ is capped at 0.5 if the number of observed events is below the number of expected events.}
  \label{table:result}
  \vspace{-0.3cm}
  \begin{center}
    {\scriptsize
      \begin{tabular}{lccc}
        \hline\noalign{\smallskip}
        & SR2A & SR2B & SR2C \\
        \hline\noalign{\smallskip}
        Data & 10 & 1 & 2 \\
        \hline\noalign{\smallskip}
        Fitted total SM & $10.8 \pm 1.7$ & $2.4\pm 0.9$ & $3.5 \pm 0.5$  \\
        $p_0$ & 0.50 & 0.50 & 0.50  \\
         \hline\noalign{\smallskip}
        Fitted $t\bar{t}$ & $7.3 \pm 1.4$ & $1.4 \pm 0.7$ & $2.4\pm 0.4$ \\
        Fitted single top & $0.61\pm 0.15$ & $0.23\pm 0.17$ & $0.10^{+0.13}_{-0.10}$\\
        Fitted $Z$+jets & $0.91\pm 0.22$ & $0.14 \pm 0.06$ & $0.16 \pm 0.06$\\
        Fitted diboson & $0.46 \pm 0.34$  & $0.27\pm 0.21$ & $0.15\pm 0.12$  \\ 
        Fitted $t\bar{t}V$, $tZ$ & $1.0 \pm 0.4$  & $0.38 \pm 0.18$  & $0.65 \pm 0.23 $   \\
        Fitted non-prompt & $0.52 \pm 0.11$ & $<0.05$  & $<0.01$   \\
        \hline\noalign{\smallskip}
        MC exp. total SM & $11.6$ & 3.0 &  4.8 \\
        \hline\noalign{\smallskip}
        MC exp. $t\bar{t}$ & 8.1 & 2.0 & 3.7  \\
        MC exp. single top & 0.61 & 0.24 & 0.14 \\
        Data-driven $Z$+jets & 0.88 & 0.13 & 0.18 \\
        MC exp. diboson & 0.48 & 0.28 & 0.15 \\
        MC exp. $t\bar{t}V$, $tZ$ & 1.0 & 0.38 & 0.66 \\
        Data-driven non-prompt & 0.52 & $<0.05$ & $<0.01$\\
        \hline\noalign{\smallskip}
      \end{tabular}
    }
    {\scriptsize
      \begin{tabular}{lcc}
        \hline\noalign{\smallskip}
        & SR3A & SR3B	\\
        \hline\noalign{\smallskip}
        Data &  4 & 2\\
        \hline\noalign{\smallskip}
        Total SM &  4.5 $\pm$ 1.4 &  1.3 $\pm$ 0.4 \\
         $p_0$ & 0.50 & 0.30 \\
        \hline\noalign{\smallskip}
        MC exp. $t\bar{t}V$, $tZ$ &  3.5 $\pm$ 1.2 & 1.1 $\pm$ 0.4 \\
        MC exp. diboson, triboson &  0.1 $\pm$ 0.1 & 0.1 $\pm$ 0.1 \\ 
        MC exp. $Wh$, $Zh$, $\ttbar h$ & 0.1 $\pm$ 0.1 & 0.04 $\pm$ 0.04 \\
        Data-driven non-prompt &  0.8 $\pm$ 0.7  & $<0.2$ \\
        \hline\noalign{\smallskip}
      \end{tabular}
    }
  \end{center}
\end{table}

\begin{figure}[!htb]
  \begin{center}
    \includegraphics[scale=0.38]{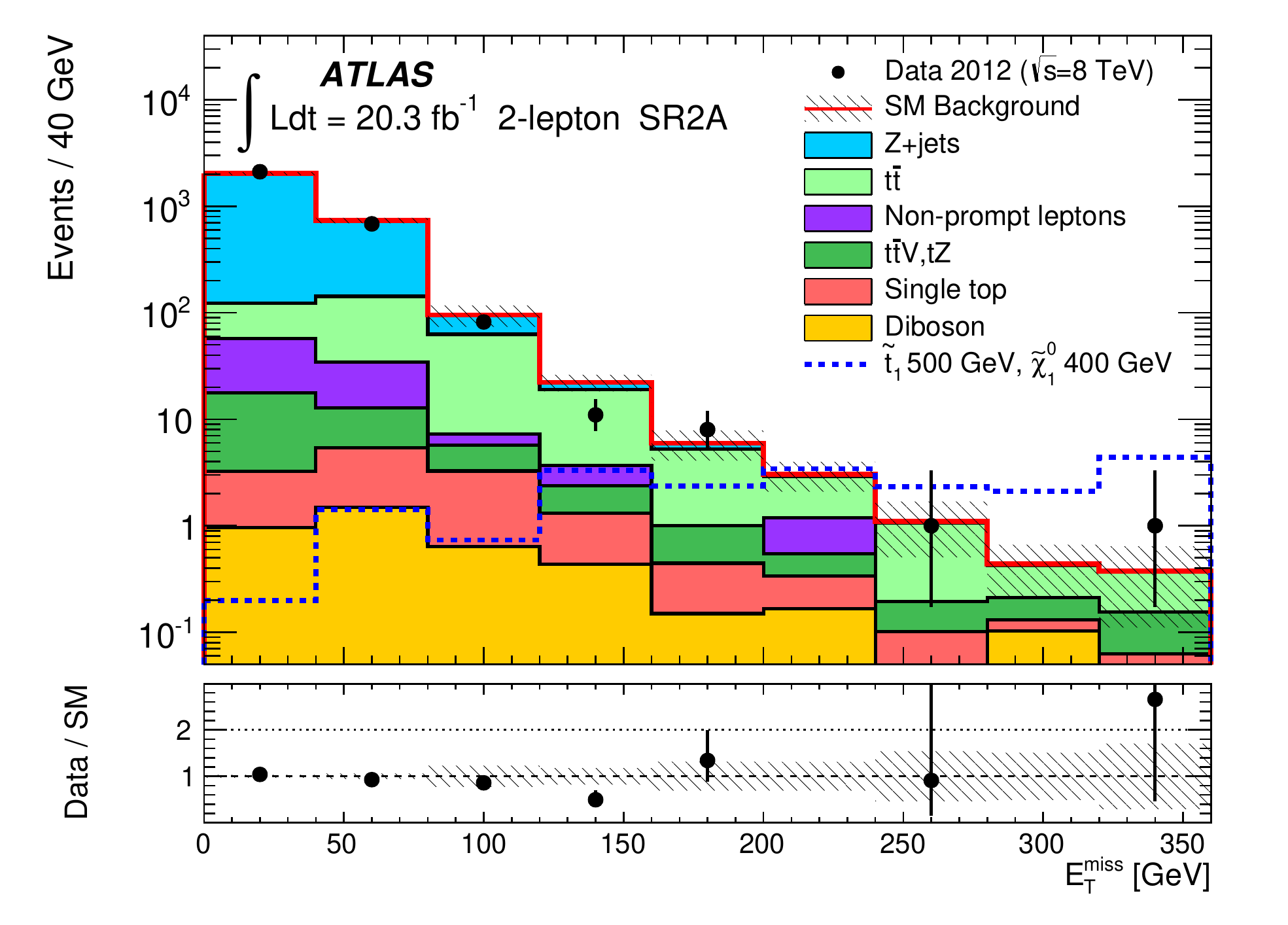}
    \includegraphics[scale=0.38]{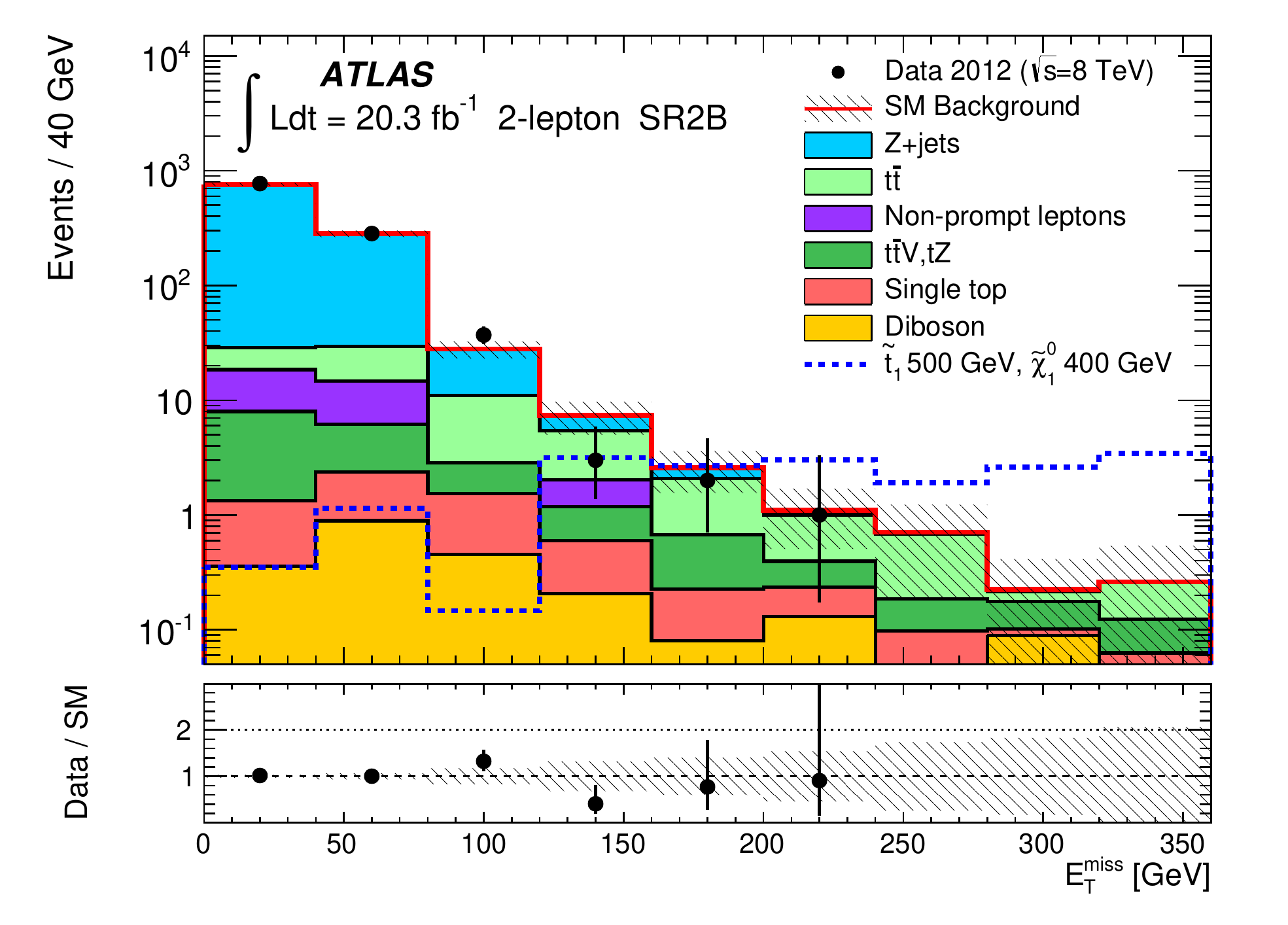}
    \includegraphics[scale=0.38]{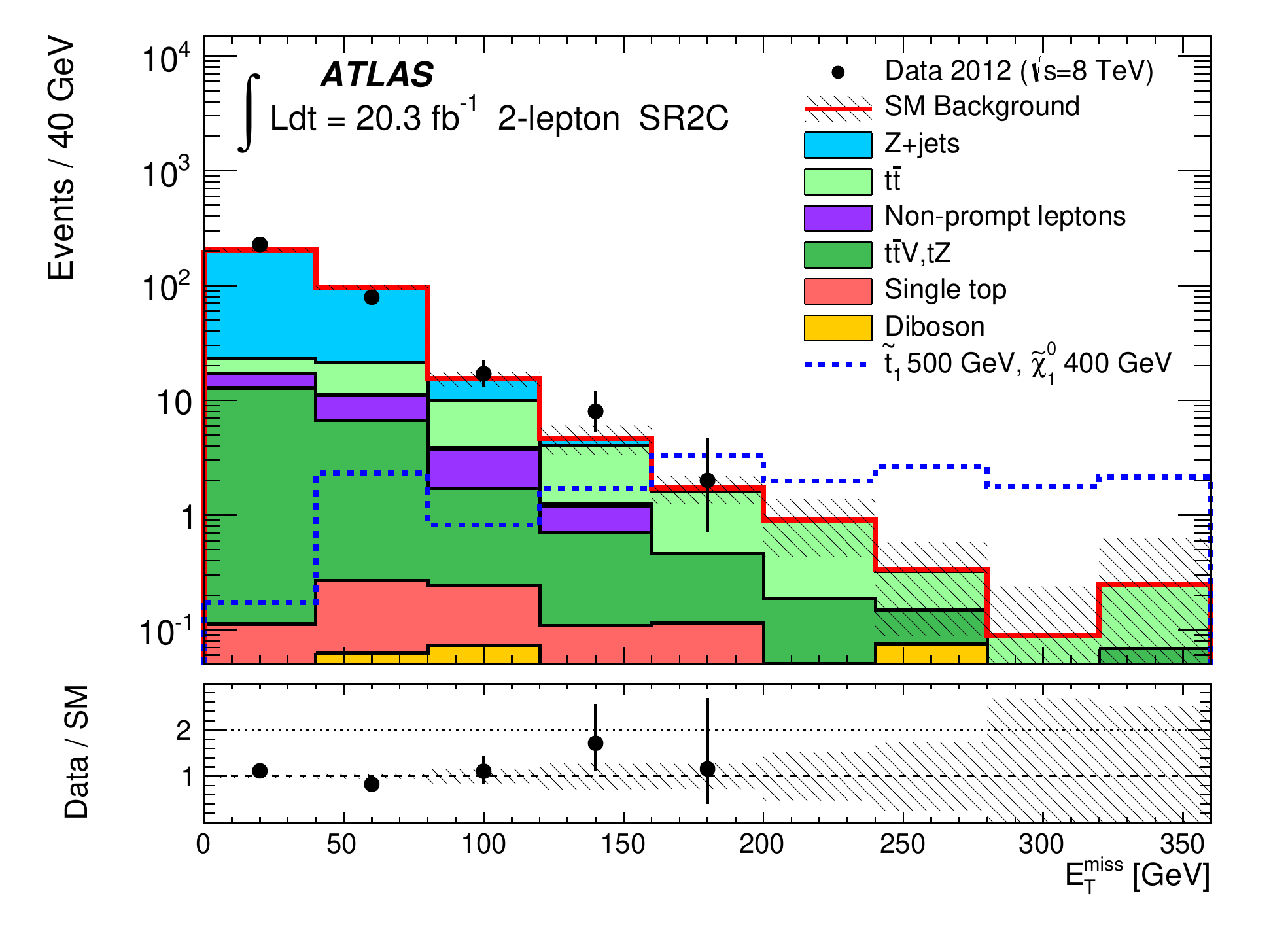}
  \end{center}
    \vspace{-0.3cm}
  \caption{The missing transverse momentum distribution for the  2-lepton SRs SR2A ({\it top}), SR2B ({\it middle}) and SR2C ({\it bottom}) 
    before the final $\met$ selection after the background fit.
    $Z$+jets distributions are obtained using the jet smearing method.
    Shaded bands denote the statistical and systematic uncertainty on the background. For illustration, distributions for a GMSB signal scenario with
    $m_{\tilde{t}_{1}}=500\GeV$, $m_{\tilde{\chi}_{1}^{0}}=400\GeV$ are shown. The last bin includes the histogram overflow.}
  \label{figure:2lep_final_met}
\end{figure}

\begin{figure}[!htb]
\begin{center}
  \subfigure{\includegraphics[scale=0.42]{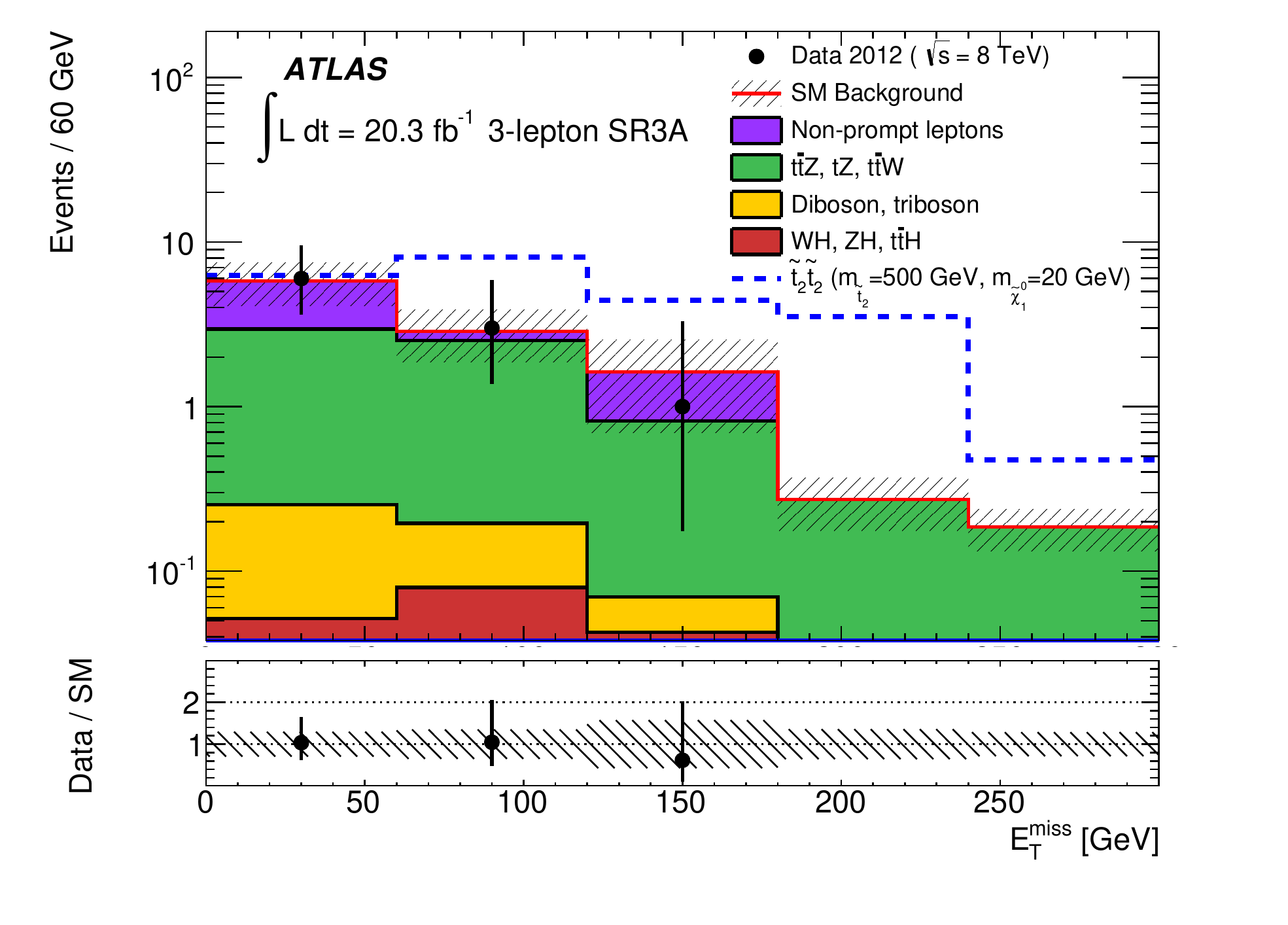}}
   \hspace{-10pt}
   \subfigure{\includegraphics[scale=0.42]{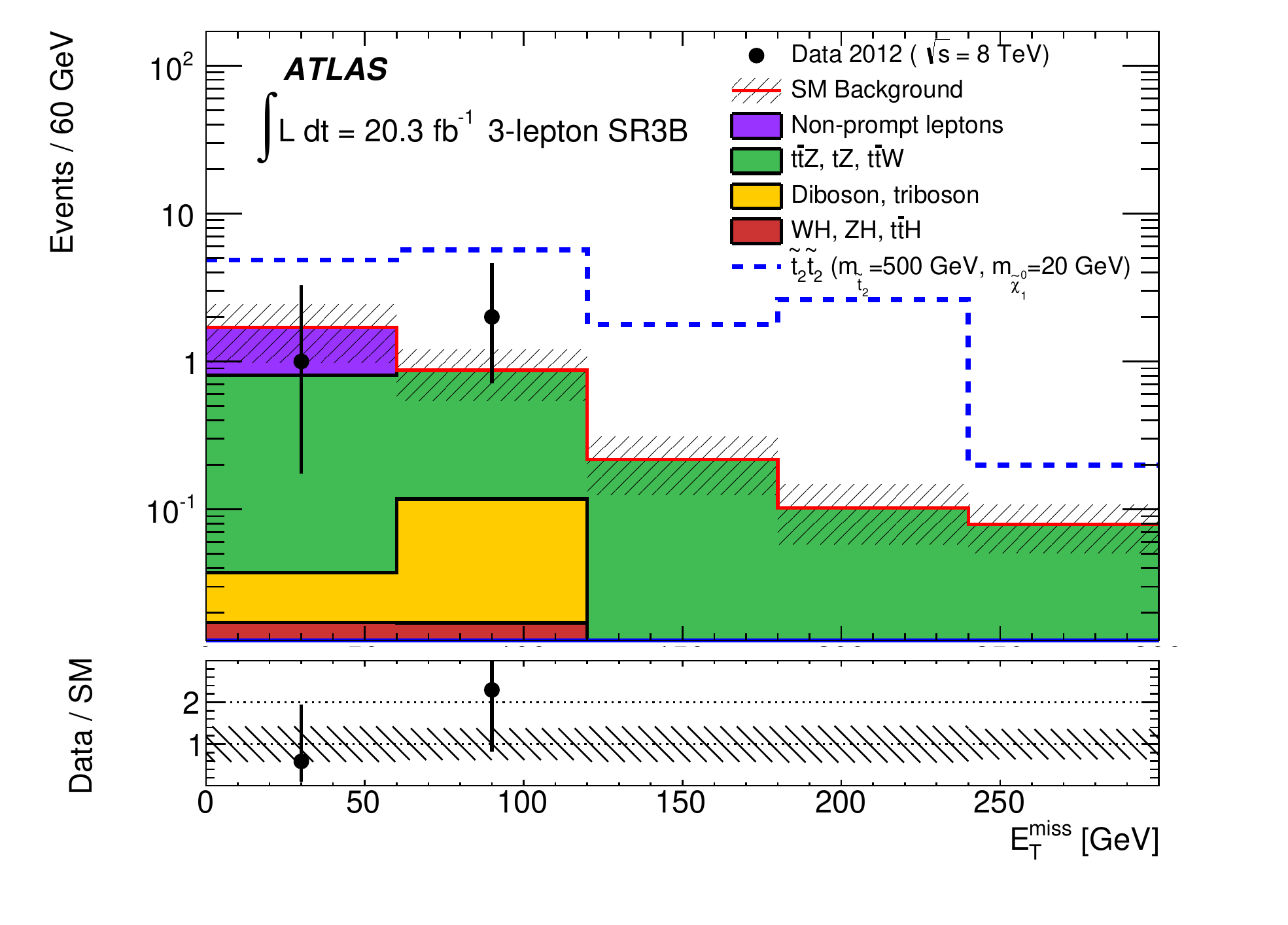}}  
   \end{center}
   \vspace{-0.7cm}
   \caption{The missing transverse momentum for the 3-lepton SRs SR3A ({\it top}) and SR3B ({\it bottom}) before the final $\met$ selection. Shaded bands denote the statistical and systematic uncertainty on the background. For illustration, distributions for a signal point in the $\ttwo$ simplified model with $m_{\ttwo}=500$~GeV and  $m_{\tilde{\chi}^{0}_{1}}=20$~GeV are also shown.
     The last bin includes the histogram overflow.}
\label{fig:3lep_met_finalSR}
\end{figure}

No excess is observed in any of the SRs. The probability ($p_0$-value) of the SM background 
to fluctuate to the observed number of events or higher in each SR is also reported in Table~\ref{table:result}, and has been truncated at 0.5. 
Upper limits at 95\% CL on the number of beyond the SM (BSM) events for each SR are derived using the CL$_s$ prescription~\cite{Read:2002hq} 
and neglecting any possible signal contamination in the control regions. After normalising these by the integrated luminosity of the data sample, they can be interpreted as upper limits on the visible BSM cross section, 
$\sigma_{\rm vis}$, defined as the product of acceptance, reconstruction efficiency and production cross section.
The limits are calculated from pseudo-experiments as well as with asymptotic formulae~\cite{AsymptoticFormula} for comparison.
The results are given in Table~\ref{table:Model_indep}.  

\begin{table}[!htb]
\caption{Signal model independent upper limits on 
  the visible signal cross section ($\sigma_{\rm{vis}}=\sigma_{\rm{prod}}\times A \times\epsilon$) in the five SRs. 
  The numbers (in parenthesis) give the observed (expected) 95\% CL upper
  limits. Calculations are performed with pseudo-experiments.
  The $\pm$1$\sigma$ variations on the expected limit due to the statistical and background systematic uncertainties are also shown. 
  The equivalent limits on the visible cross section calculated using an asymptotic
    method are given inside the square brackets.
}
\label{table:Model_indep}
\begin{center}
\renewcommand{\arraystretch}{1.15}
  \begin{tabular}{lll}
    \hline\noalign{\smallskip}
    Signal region & \multicolumn{2}{c}{$\sigma_{\rm{vis}}$ [fb]} \\ 
    \hline\noalign{\smallskip}
    SR2A & 0.40 ($0.46^{+0.16}_{-0.13}$) & [0.39 ($0.41^{+0.20}_{-0.12}$)]\\
    SR2B & 0.19 ($0.24^{+0.07}_{-0.05}$) & [0.19 ($0.22^{+0.13}_{-0.05}$)] \\
    SR2C  & 0.20 ($0.27^{+0.11}_{-0.07}$) & [0.20 ($0.27^{+0.13}_{-0.08}$)] \\
    SR3A &  0.30 ($0.31^{+0.14}_{-0.05}$) & [0.29 ($0.31^{+0.16}_{-0.10}$)] \\
    SR3B &  0.26 ($0.20^{+0.08}_{-0.02}$) & [0.24 ($0.20^{+0.11}_{-0.05}$)] \\
    \hline\noalign{\smallskip}
  \end{tabular}
\end{center}
\end{table}

These results are also 
interpreted in the context of the models described in Section~\ref{Introduction}. 
Exclusion limits are calculated by combining the results from several exclusive SRs. 
For the GMSB scenarios, SR2C and SR3A are combined with the region with best expected sensitivity between SR2A or SR2B. 
For the $\ttwo$ simplified models, SR2C is combined with the region with best expected sensitivity between SR3A or SR3B. 
For model-dependent interpretations, the fit described in Section~\ref{section:background_estimation} is modified to
include the expected signal contamination of the CRs and the observed number of events in the SRs as well as an extra free parameter for a possible BSM signal strength which is constrained to be non-negative. 
The expected and observed exclusion limits are calculated using asymptotic formulae for each SUSY model point, 
taking into account the theoretical and experimental uncertainties on the SM background and the experimental uncertainties on the signal. 
The impact of the uncertainties on the signal cross section is also addressed for the observed limit only 
by showing the results obtained when moving the nominal 
cross section up or down by the $\pm 1\sigma$ theoretical uncertainty. 
Quoted numerical limits on the particle masses refer to the signal cross sections reduced by 1$\sigma$. 

\begin{figure}[t]
  \begin{center}
    \includegraphics[scale=0.42]{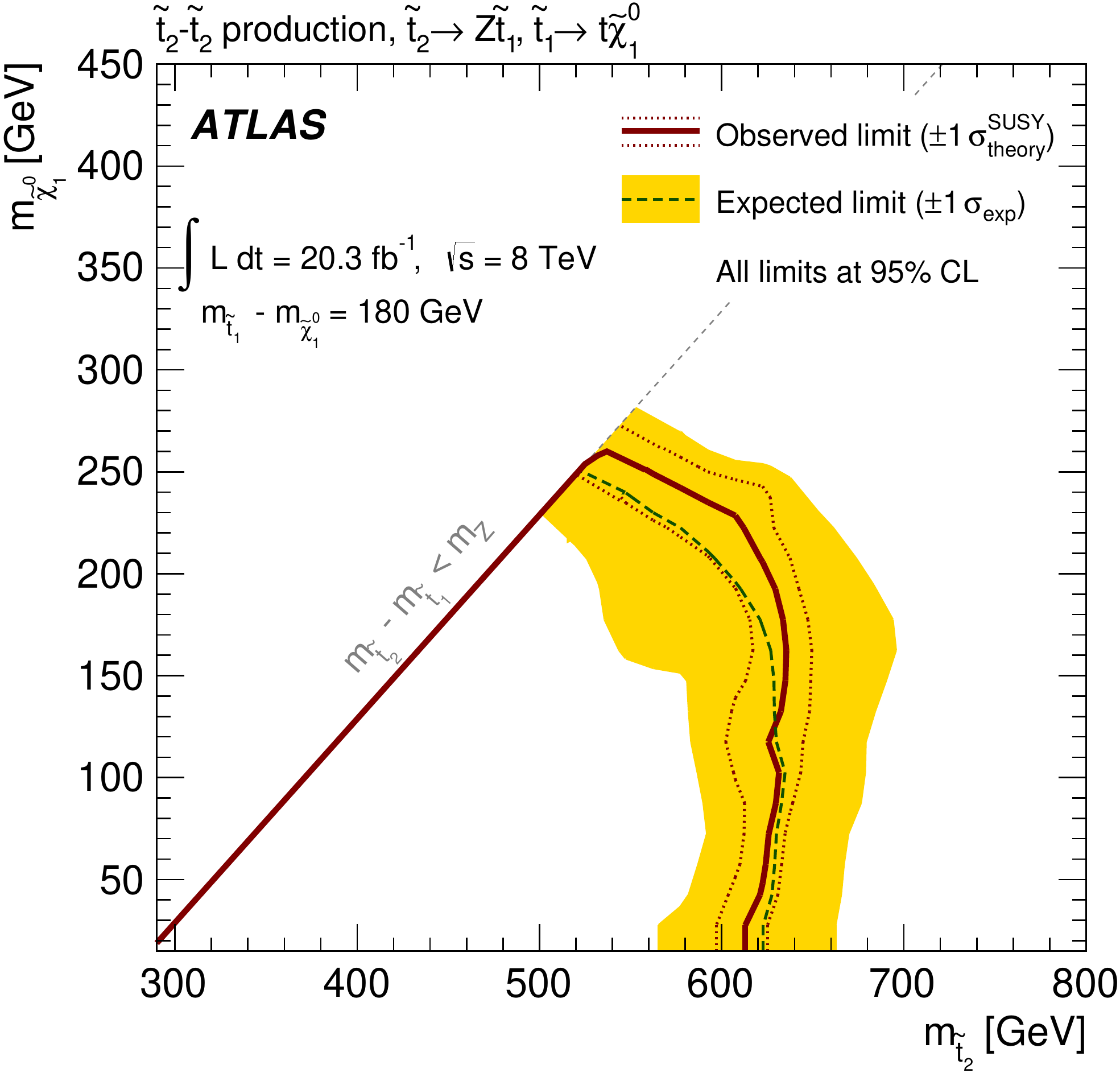}
  \end{center}
  \vspace{-0.5cm}
  \caption{Expected and observed exclusion limits in the $m_{\ttwo}$-$m_{\tilde{\chi}_1^0}$ plane 
  for the direct $\ttwo$ pair production simplified model with $\text{BR}(\ttwo\to Z\tone)=1$. 
  The contours of the band around the expected limit are the $\pm$1$\sigma$ results, 
  including all uncertainties except theoretical uncertainties on the signal cross section. The dotted lines around the observed
    limit illustrate the change in the observed limit as the nominal signal cross section is scaled up and down
    by the theoretical uncertainty. All limits are computed at 95\% CL. }
  \label{fig:3lep_limitSRComb}
\end{figure}

\begin{figure*}[t]
  \begin{center}
    \includegraphics[scale=0.7]{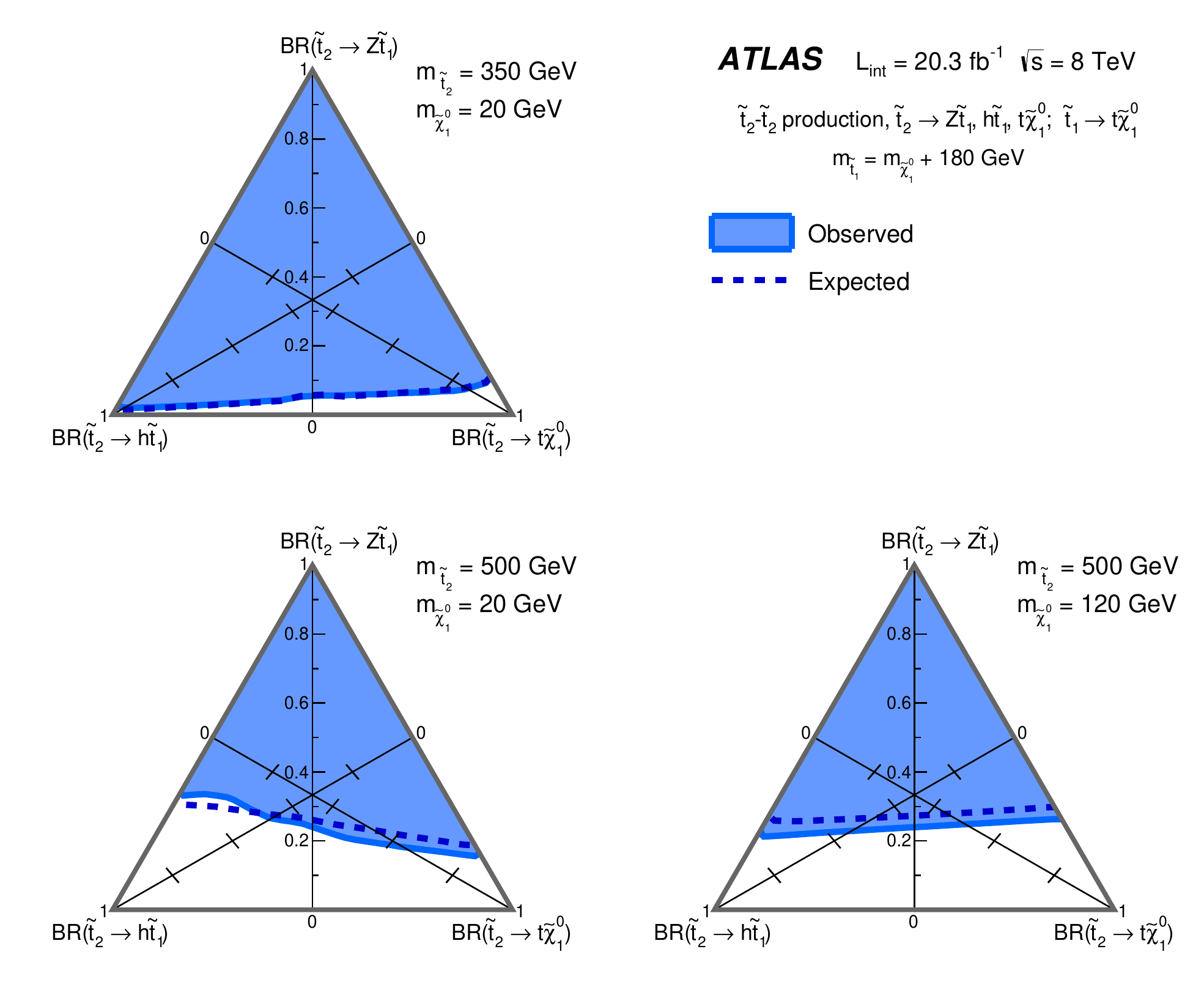}
  \end{center}
  \vspace{-0.4cm}
  \caption{
    Exclusion limits at 95\% CL are shown for the direct $\ttwo$ pair production simplified model  
    as a function of the branching ratios $\text{BR}(\ttwo\to Z\tone)$, $\text{BR}(\ttwo\to h\tone)$ and $\text{BR}(\ttwo\to t\tilde{\chi}_1^0$) 
    for $(m_{\ttwo},m_{\tilde{\chi}_1^0})=(350,20) \GeV$ ({\it top}), $(500,20) \GeV$ ({\it bottom left}) and $(500,120)\GeV$ ({\it bottom right}). 
    The dashed and solid lines show the expected and observed limits, respectively, including all uncertainties except the theoretical signal cross section uncertainty (PDF and scale). }
  \label{fig:3lep_limitSRComb_ternary}
\end{figure*}

Figure~\ref{fig:3lep_limitSRComb} shows the limit obtained in the $\ttwo$ simplified model, 
which excludes 
$m_{\ttwo}<525\GeV$ for
$m_{\tilde{\chi}_1^0}<240\GeV$ and  $m_{\ttwo}<600\GeV$ for
$m_{\tilde{\chi}_1^0}<200\GeV$. 
The interpolation of the limit contours between the simulated points towards the $\ttwo\to Z\tone$ kinematic boundary 
has been established using MC generator level information. 
A reduction in acceptance of up to 20\% is observed in the region where 
$m_{\ttwo}-m_{\tone}-m_{Z}$ is comparable to the $Z$ boson width. 
The region with $m_{\ttwo}-m_{\tone}<m_{Z}$,  where the $\ttwo\to Z^{(*)}\tone$ decay involves an off-shell $Z$, has not been considered since in that case 
other $\ttwo$ decay modes, such as $\ttwo\to t\tilde{\chi}_{1}^{0}$, would be dominant.
If the assumption on the 100\% branching ratio for the $\ttwo\to Z \tone$ decay mode is relaxed,
the $\ttwo$ can also decay via $\ttwo\to h\tone$ and $\ttwo\to t\tilde{\chi}_{1}^{0}$. 
Exclusion limits as a function of the $\ttwo$ branching ratios are shown in Figure~\ref{fig:3lep_limitSRComb_ternary} 
for representative values of the masses of $\ttwo$ and $\tilde{\chi}_1^0$. 
For low $\ttwo$ mass ($m_{\ttwo}=350\GeV$), SUSY models with $\text{BR}(\ttwo\to Z\tone)$ 
above 10\% are excluded. For higher stop mass ($m_{\ttwo}=500\GeV$), models
with $\text{BR}(\ttwo\to Z\tone)$ above 15-30\% are excluded, with a small dependence on the value
of the neutralino mass, $\text{BR}(\ttwo\to h\tone)$ and $\text{BR}(\ttwo\to t\tilde{\chi}_{1}^{0})$.

In Figure~\ref{figure:GMSB_limit} the expected and observed limits are shown for the GMSB scenarios on the $\tone$, $\tilde{\chi}_1^0$ mass plane. 
Stop masses up to $540\GeV$ are excluded for neutralino masses of $100 \GeV<m_{\tilde{\chi}_{1}^{0}}<m_{\tilde{t}_{1}}-10\GeV$. 
In the parameter space region where the $\tone$ only decays via $b\tilde{\chi}_{1}^{\pm}$,
the exclusion extends up to stop masses of $660\GeV$ for neutralinos of $550\GeV$. 
For illustration, the exclusion limits
obtained with 2.05~$\ifb$ of ATLAS data at $\sqrt{s}=7\TeV$ for the similar model are also shown, in which
the maximum limit on the stop masses was $330\GeV$. Due to the increase in statistics and the proton-proton collision energy,
as well as the optimised selections for these conditions, much stronger constraints are now set on this model.

\begin{figure}[!htb]
  \begin{center}
    \includegraphics[scale=0.42]{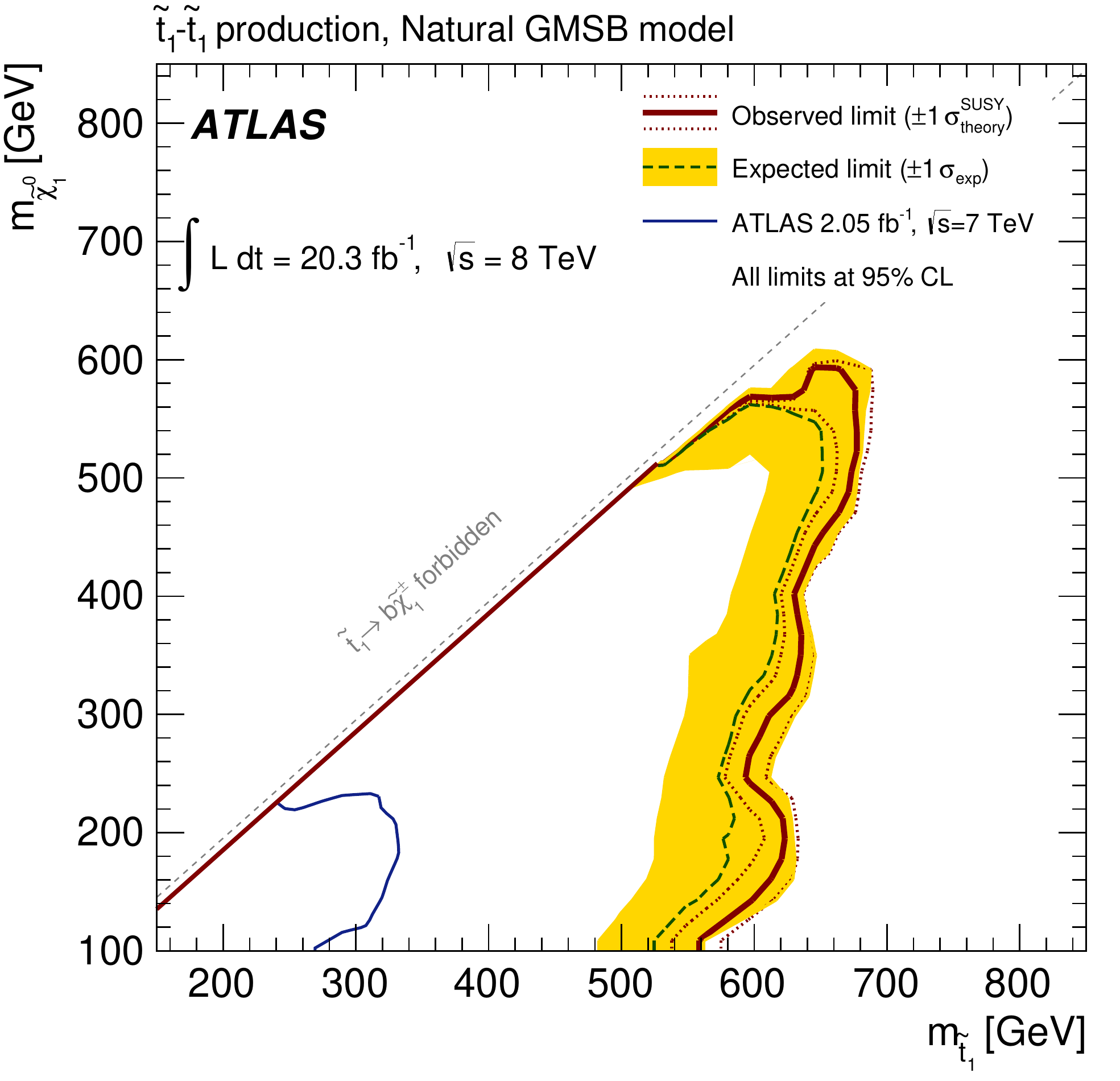}
  \end{center}
    \vspace{-0.4cm}
  \caption{Expected and observed exclusion limits at 95\% CL for the stop natural GMSB model described in the text. 
    The contours of
    the band around the expected limit are the $\pm$1$\sigma$ results, 
    including all uncertainties except theoretical uncertainties on the signal cross section. The dotted lines around the observed
    limit illustrate the change in the observed limit as the nominal signal cross section is scaled up and down
    by the theoretical uncertainty. For comparison, the observed exclusion
    limit with 2.05~fb$^{-1}$ of data at $\sqrt{s}=7\TeV$ at ATLAS for a similar model~\cite{AtlasStopGMSB2011} is shown.}
  \label{figure:GMSB_limit}
\end{figure}

\section{Summary and Conclusions}

This paper presents a dedicated search for direct stop pair production
in decays with an experimental signature compatible
with the production of a $Z$ boson, {\it b}-jets and missing transverse momentum. 
The analysis is performed with $pp$ collision data at $\sqrt{s}=8\TeV$ collected with the ATLAS detector at the LHC   
corresponding to an integrated luminosity of 20.3~fb$^{-1}$. 
The results are interpreted in the framework of 
simplified models with production of $\ttwo$ as well as in a natural GMSB model.

In a simplified model characterised by the decay chain $\ttwo\to Z\tone$ with $\tone\to t\tilde{\chi}_{1}^{0}$ and 
the mass difference between $\tone$ and $\tilde{\chi}_{1}^{0}$ slightly larger than the top mass,
parameter space regions with 
$m_{\ttwo}<600\GeV$ and $m_{\tilde{\chi}_1^0}<200 \GeV$ are excluded at 95\% CL. 
When the $\ttwo\to h\tone$ and $\ttwo\to t\tilde{\chi}_1^0$ decays 
are included in the model, $\text{BR}(\ttwo\to Z\tone)>10{\text -}30\%$ are excluded for several mass configurations. 
These are the first experimental results on the search for $\ttwo$.

In the GMSB scenario, where the $\tone$ might decay to $b\tilde{\chi}_{1}^{\pm}$ or $t\tilde{\chi}_{1}^{0}(\tilde{\chi}_{2}^{0})$
and the $\tilde{\chi}_{1}^{0}$ decay in $Z\tilde{G}$ or $h\tilde{G}$, parameter space regions with $\tone$ masses below $540\GeV$ are
excluded at 95\% CL for $100\GeV < m_{\tilde{\chi}_{1}^{0}}<m_{\tilde{t}_{1}}-10\GeV$. 
These limits are much stronger than those set on the similar model considered in the search at $\sqrt{s}=7\TeV$.
For $\tilde{\chi}_{1}^{0}$ masses of about $550\GeV$, better sensitivity is achieved and $\tone$ masses below $660\GeV$ are excluded.

\begin{acknowledgements}
\input{Acknowledgements.tex}

\end{acknowledgements}

\bibliographystyle{atlasBibStyleWoTitle}     
\bibliography{PAPER-stopZ-2013}

\onecolumn 
\clearpage
\input{atlas_authlist} 

\end{document}

%% file: Acknowledgements.tex


We thank CERN for the very successful operation of the LHC, as well as the
support staff from our institutions without whom ATLAS could not be
operated efficiently.

We acknowledge the support of ANPCyT, Argentina; YerPhI, Armenia; ARC,
Australia; BMWF and FWF, Austria; ANAS, Azerbaijan; SSTC, Belarus; CNPq and FAPESP,
Brazil; NSERC, NRC and CFI, Canada; CERN; CONICYT, Chile; CAS, MOST and NSFC,
China; COLCIENCIAS, Colombia; MSMT CR, MPO CR and VSC CR, Czech Republic;
DNRF, DNSRC and Lundbeck Foundation, Denmark; EPLANET, ERC and NSRF, European Union;
IN2P3-CNRS, CEA-DSM/IRFU, France; GNSF, Georgia; BMBF, DFG, HGF, MPG and AvH
Foundation, Germany; GSRT and NSRF, Greece; ISF, MINERVA, GIF, I-CORE and Benoziyo Center,
Israel; INFN, Italy; MEXT and JSPS, Japan; CNRST, Morocco; FOM and NWO,
Netherlands; BRF and RCN, Norway; MNiSW and NCN, Poland; GRICES and FCT, Portugal; MNE/IFA, Romania; MES of Russia and ROSATOM, Russian Federation; JINR; MSTD,
Serbia; MSSR, Slovakia; ARRS and MIZ\v{S}, Slovenia; DST/NRF, South Africa;
MINECO, Spain; SRC and Wallenberg Foundation, Sweden; SER, SNSF and Cantons of
Bern and Geneva, Switzerland; NSC, Taiwan; TAEK, Turkey; STFC, the Royal
Society and Leverhulme Trust, United Kingdom; DOE and NSF, United States of
America.

The crucial computing support from all WLCG partners is acknowledged
gratefully, in particular from CERN and the ATLAS Tier-1 facilities at
TRIUMF (Canada), NDGF (Denmark, Norway, Sweden), CC-IN2P3 (France),
KIT/GridKA (Germany), INFN-CNAF (Italy), NL-T1 (Netherlands), PIC (Spain),
ASGC (Taiwan), RAL (UK) and BNL (USA) and in the Tier-2 facilities
worldwide.

%% file: atlas_authlist.tex
\begin{flushleft}
{\Large The ATLAS Collaboration}

\bigskip

G.~Aad$^{\rm 84}$,
B.~Abbott$^{\rm 112}$,
J.~Abdallah$^{\rm 152}$,
S.~Abdel~Khalek$^{\rm 116}$,
O.~Abdinov$^{\rm 11}$,
R.~Aben$^{\rm 106}$,
B.~Abi$^{\rm 113}$,
M.~Abolins$^{\rm 89}$,
O.S.~AbouZeid$^{\rm 159}$,
H.~Abramowicz$^{\rm 154}$
% ...
\bigskip
\\
% ...
$^{11}$ Institute of Physics, Azerbaijan Academy of Sciences, Baku, Azerbaijan\\
% ...
$^{84}$ CPPM, Aix-Marseille Universit{\'e} and CNRS/IN2P3, Marseille, France\\
% ...
$^{89}$ Department of Physics and Astronomy, Michigan State University, East Lansing MI, United States of America\\
% ...
$^{106}$ Nikhef National Institute for Subatomic Physics and University of Amsterdam, Amsterdam, Netherlands\\
% ...
$^{112}$ Homer L. Dodge Department of Physics and Astronomy, University of Oklahoma, Norman OK, United States of America\\
$^{113}$ Department of Physics, Oklahoma State University, Stillwater OK, United States of America\\
% ...
$^{116}$ LAL, Universit{\'e} Paris-Sud and CNRS/IN2P3, Orsay, France\\
% ...
$^{152}$ Institute of Physics, Academia Sinica, Taipei, Taiwan\\
% ...
$^{154}$ Raymond and Beverly Sackler School of Physics and Astronomy, Tel Aviv University, Tel Aviv, Israel\\
% ...
$^{159}$ Department of Physics, University of Toronto, Toronto ON, Canada\\
% ...
\end{flushleft}
